\documentstyle[12pt,epsf]{article}

\renewcommand{\thefootnote}{\fnsymbol{footnote}}

\newcommand{\lmm}{\ln\frac{\mu^2}{m^2}}
\newcommand{\lmmz}{\ln^2\frac{\mu^2}{m^2}}
\newcommand{\Lmm}{\ln\frac{\mu^2}{m^2}}
\newcommand{\Lmmz}{\ln^2\frac{\mu^2}{m^2}}

\begin{document}    

\begin{titlepage}
\noindent                                          
%
% Date
%
\begin{flushright}
{\bf \hfill TTP96--13}\footnote{The complete postscript file of this
preprint, including figures, is available via anonymous ftp at
www-ttp.physik.uni-karlsruhe.de (129.13.102.139) as /ttp96-13/ttp96-13.ps 
or via www at http://www-ttp.physik.uni-karlsruhe.de/cgi-bin/preprints.}\\
{\bf MPI/PhT/96-27}\\
{\bf \hfill hep-ph/9606230}\\
{\bf \hfill May 1996}
\end{flushright}

\vspace{.5cm}

\begin{center}
\begin{LARGE}
\mathversion{bold}
 {\bf 
Three-Loop Polarization Function and\\ 
${\cal O}(\alpha_s^2)$ Corrections to the Production\\
 of Heavy Quarks\footnote{
     Work supported by BMFT under Contract 056KA93P6, 
     DFG under Contract Ku502/6-1 and INTAS under Contract INTAS-93-0744.}
 }
\mathversion{normal}
\end{LARGE}

\vspace{.8cm}

\begin{large}
 K.G.~Chetyrkin$^{a}$,
 J.H.~K\"uhn$^{b}$, 
 M.~Steinhauser$^{b}$\footnote{Supported by 
              Graduiertenkolleg Elementarteilchenphysik, Karlsruhe.}
\end{large}

\vspace{.4cm}

\begin{itemize}\centering
\item[$^a$]{\it
   Max-Planck-Institut f\"ur Physik, Werner-Heisenberg-Institut,\\
   D-80805 Munich, Germany\\
   and\\
   Institute for Nuclear Research, Russian Academy of Sciences,\\
   Moscow 117312, Russia
           }
\vspace{3mm}
\item[$^b$]{\it
   Institut f\"ur Theoretische Teilchenphysik, Universit\"at Karlsruhe,\\
   D-76128 Karlsruhe, Germany.\\  
           }
\end{itemize}

\vspace{.5cm}

\begin{abstract}
\noindent
The three-loop vacuum polarization function $\Pi(q^2)$
induced by a massive quark is calculated. A comprehensive
description of the method is presented.
From the imaginary part the ${\cal O}(\alpha_s^2)$ result
for the production of heavy quarks 
$R(s)=\sigma(e^+e^-\to \mbox{hadrons})/\sigma(e^+e^-\to \mu^+\mu^-)$
can be extracted. Explicit formulae separated into the different
colour factors are given.
\end{abstract}

%\noindent
%emails:
%\\
%ms@ttpux2.physik.uni-karlsruhe.de\\

\vfill
\end{center}
\end{titlepage}

%%%%%%%%%%%%%%%%%%%%%%%%%%%%%%%%%%%%%%%%%%%%%%%%%%%%%%%%%%%%
%%%%%%%%%%%%%%%%%%%%%%%%%%%%%%%%%%%%%%%%%%%%%%%%%%%%%%%%%%%%

\renewcommand{\thefootnote}{\arabic{footnote}}
\setcounter{footnote}{0}

\section{Introduction}

The measurement of the total cross section for
electron positron annihilation into hadrons allows for
a unique test of pertubative QCD. The decay rate
$\Gamma(Z \to \mbox{hadrons})$ provides one of the
most precise determinations of the strong coupling 
constant $\alpha_s$. In the high energy limit the quark masses
can often be neglected.
In this approximation QCD corrections to 
$R  \equiv \sigma(e^+ e^- \to \mbox{hadrons})/
           \sigma(e^+ e^- \to \mu^+ \mu^-)$
have been calculated up to order $\alpha_s^3$
\cite{CheKatTka79DinSap79CelGon80,GorKatLar91,SurSam91}.
For precision measurements the dominant mass corrections must be
included through an expansion in $m^2/s$. Terms up to order 
$\alpha_s^3 m^2/s$ 
\cite{CheKue90}
and $\alpha_s^2 m^4/s^2$ 
\cite{CheKue94}
are available at present,
providing an acceptable approximation from the high energy
region down to intermediate energy values.
For a number of measurements, however, the information on the 
complete mass dependence is desirable. This
includes charm and bottom meson production above the resonance 
region, say
above $4.5$~GeV and $12$~GeV, respectively, and of course
top quark production at a future electron positron collider.

To order $\alpha_s$ this calculation was performed by
K\"all\'en and Sabry in the context of QED a long time ago
\cite{KaeSab55}.
With measurements of ever increasing precision, predictions
in order $\alpha_s^2$ are needed for a reliable
comparison between theory and experiment. Furthermore, 
when one tries to apply the ${\cal O}(\alpha)$ result 
to QCD, with its running coupling
constant, the choice of scale becomes important.
In fact, the distinction between the two intrinsically different 
scales, the relative momentum versus the center of mass
energy, is crucial for a stable numerical prediction.
This information can be obtained from a full calculation 
to order $\alpha_s^2$ only. 
Such a calculation then allows to predict the cross section 
in the complete energy region where perturbative QCD can be applied
--- from close to threshold up to high energies.

The two-loop result had been calculated  in analytical form
\cite{KaeSab55}
in terms of polylogarithms. The strategy employed in
\cite{KaeSab55}
was based on a direct evaluation of the imaginary part
of the vacuum polarization (and thus of the cross section)
in a first step followed by the calculation of the
real part via dispersion relations. In fact, this strategy
was also applied in the two-loop calculation of the polarization function
$\Pi$ for the unequal mass case which enters e.g. the 
$\rho$ parameter
\cite{Djo88,KniKueStu88,Kni90}.

The same holds true for a special class of three-loop contributions 
\cite{HoaKueTeu95}
to $\Pi$ 
which are present in QED and QCD 
as well --- those induced by massless fermion loop insertions into
the internal gluon propagator.
The imaginary part was calculated in terms of polylogarithms with
a method fairly similar to the one employed in order $\alpha$,
with infrared cutoffs introduced to calculate separately
real and virtual contributions.
However, the calculation techniques of 
\cite{HoaKueTeu95}
were tailored to the ``double bubble'' topology and cannot be generalized
to those which arise for
purely photonic or gluonic corrections.

A complete different approach has been employed for example in the
original calculations of the ${\cal O}(\alpha_s^2)$
corrections to $R$ for massless quarks: The full analytical function
$\Pi(q^2)$ is calculated first, and the imaginary part is obtained as a 
trivial by-product. The pitfalls from infrared divergencies due to 
the separation into real and virtual radiation are thus circumvented.
However, the Feynman integrals to be solved are generally not
available in closed analytical form once the quark masses
are different from zero. We will demonstrate in this paper that
this drawback can be circumvented by numerical methods which
allow to calculate real and imaginary part simultaneously.
Those methods exploit again heavily the analyticity of
$\Pi(q^2)$. Important ingredients are the leading terms of the expansion in
the high energy region $(-q^2)/m^2\gg 1$, the Taylor series around $q^2=0$
which has been evaluated up to terms of order $(q^2)^7$
and information about the structure of $\Pi$ in the threshold region.
The method can be tested in the case of the ``double bubble''
diagrams against the known analytical result with highly satisfactory
results. Increasing the number of terms in the large and small $q^2$
region, the $\Pi(q^2)$ can be reconstructed with arbitrary
precision. However, the results presented below should be sufficiently
precise for comparison with experimental results in the forseable
future. 

The contributions $\sim C_F^2, \sim C_A C_F$ and $\sim C_F T n_l$
have to be treated separately since they differ significantly
in their  singularity structure.
For each of the three functions an interpolation is constructed
which incorporates all data and is based on conformal
mapping and Pad\'e approximation suggested in
\cite{FleTar94,BroFleTar93,BroBaiIly94,BaiBro95}. 
Since the result for $C_F T n_l$ is available in closed form the
approximation method can be tested and shown to give excellent 
result for this case. Reliable predictions for $R$ to order
$\alpha_s^2$ and arbitrary $m^2/s$ are thus available.

In this paper only results without renormalization group improvement 
and resummation of the Coulomb singularities from higher orders  
are presented. Resummation of leading higher order terms, phenomenological
applications and a more detailed discussion of our methods will be
presented elsewhere.

This work decribes the used method in detail and extends
the analysis from 
\cite{CheKueSte95Pade}, 
where mainly results were presented.
It is organized as follows: The notation and a brief
layout of our methods are presented in Section
\ref{secnot}.
The large $q^2$ behaviour of $\Pi$ is listed in Section
\ref{seclar},
the ansatz for the threshold singularities in Section
\ref{secthr}.
Most of the calculational efforts were spent on the evaluation of the 
seven lowest Taylor coefficients of $\Pi(q^2)$ presented in Section 
\ref{secsma}.
As a simple application of the result for the first Taylor
coefficient the contribution of a massive quark to the 
relation between the QED coupling constant in the 
$\overline{\mbox{MS}}$ and the on-shell scheme is presented
in Section
\ref{secQED}.
An approximation which makes use of the analyticity structure
of $\Pi$ and the results of Section 
\ref{seclar} -- \ref{secsma} is constructed in Section 
\ref{secoptappr}. The numerical results are presented 
in Section 
\ref{secres}, 
together with compact representations in terms
of simple functions. Section 
\ref{seccon} contains a brief summary and conclusions.

\section{\label{secnot}Notation and Layout of the Calculation}

In the present approach, originally suggested in
\cite{BaiBro95},
one calculates real and imaginary parts of the vacuum polarization 
$\Pi(q^2)$ and exploits heavily its analyticity properties.
The physical observable $R(s)$ is related to $\Pi(q^2)$ by
\begin{eqnarray}
R(s)   &=&  12\pi\, \mbox{Im}\Pi(q^2+i\epsilon).
\label{rtopi}
\end{eqnarray}

It is convenient to define
\begin{equation}
\Pi(q^2) = \Pi^{(0)}(q^2) 
         + \frac{\alpha_s(\mu^2)}{\pi} C_F \Pi^{(1)}(q^2)
         + \left(\frac{\alpha_s(\mu^2)}{\pi}\right)^2\Pi^{(2)}(q^2)
         + \cdots
\end{equation}
with the $\overline{\mbox{MS}}$ coupling related to 
$\Lambda_{\tiny\overline{\mbox{MS}}}$
through
\begin{eqnarray}
{\alpha_s(\mu^2)\over\pi}&=&{1\over\beta_0\ln(\mu^2/
                          \Lambda_{\tiny\overline{\mbox{MS}}}^2)}
                          \left[1-{\beta_1\over\beta_0^2}\,
                {\ln\ln(\mu^2/\Lambda_{\tiny\overline{\mbox{MS}}}^2)
                 \over\ln(\mu^2/\Lambda_{\tiny\overline{\mbox{MS}}}^2)}
                         \right],
\\
\beta_0 &=& \frac{11}{12}C_A - \frac{1}{3}T N_f,
\\
\beta_1 &=& \frac{34}{48}C_A^2 - \frac{1}{4}C_F T N_f 
          - \frac{5}{12}C_A T N_f,
\\
 && C_F=\frac{4}{3},\,\, C_A=3,\,\, T=\frac{1}{2}.
\end{eqnarray}
The two-loop coefficient $\beta_1$ was calculated in \cite{beta1}.

The number of fermions is denoted by $N_f$ and has to be distinguished
from the number of light ($\equiv$ massless) fermions $n_l=N_f-1$.
The pole mass of the heavy fermion $m$ is related to the 
$\overline{\mbox{MS}}$ mass $\bar{m}(\mu^2)$ by
\begin{eqnarray}
\frac{\bar{m}(\mu^2)}{m} &=& 1 
+ \frac{\alpha_s(\mu^2)}{\pi}C_F\left(-1-\frac{3}{4}\Lmm\right)
+ \left(\frac{\alpha_s(\mu^2)}{\pi}\right)^2 \Bigg[
  C_F T \left(
            \frac{3}{4}
          - \frac{3}{2}\zeta(2)
        \right)
\nonumber\\
&&
+C_F^2 \left(
            \frac{7}{128}
          - \frac{15}{8}\zeta(2)
          - \frac{3}{4}\zeta(3)
          + \frac{21}{32}\Lmm
          + \frac{9}{32}\Lmmz
          + 3\zeta(2)\ln2
        \right)
\nonumber\\
&&
+ C_A C_F \left(
          - \frac{1111}{384}
          + \frac{1}{2}\zeta(2)
          + \frac{3}{8}\zeta(3)
          - \frac{185}{96}\Lmm
          - \frac{11}{32}\Lmmz
          - \frac{3}{2}\zeta(2)\ln2
          \right)
\nonumber\\
&&
+ C_F T N_f \left(
            \frac{71}{96}
          + \frac{1}{2}\zeta(2)
          + \frac{13}{24}\Lmm
          + \frac{1}{8}\Lmmz
            \right)
\Bigg],
\label{msmass2osmass}
\end{eqnarray}
with $L=\ln\mu^2/m^2$.
To describe the singularity structure of $\Pi$ in the region close to 
threshold the perturbative QCD potential
\cite{Fis77}
\begin{eqnarray}
 V_{QCD}(\vec{q}\,^2) &=& 
         -4\pi C_F\frac{\alpha_V(\vec{q}\,^2)}{\vec{q}\,^2},
\\
 \alpha_V(\vec{q}\,^2) &=&  \alpha_s(\mu^2)\Bigg[
      1 + \frac{\alpha_s(\mu^2)}{4\pi}\left(
          \left(\frac{11}{3}C_A-\frac{4}{3}T n_l\right)
          \left(-\ln\frac{\vec{q}\,^2}{\mu^2}+\frac{5}{3}\right)
          -\frac{8}{3}C_A            \right)         
\label{alphav}
\Bigg]
\end{eqnarray}
will become important, where the $C_A C_F$ and $C_F T n_l$
contributions have been displayed separately.
To transform the results from QCD to QED the proper group 
theoretical coefficients $C_F=1, C_A=0$ and $T=1$ have to be 
used.

\begin{figure}[ht]
 \begin{center}
 \begin{tabular}{ccc}
   \epsfxsize=2.5cm
   \leavevmode
   \epsffile[170 280 430 520]{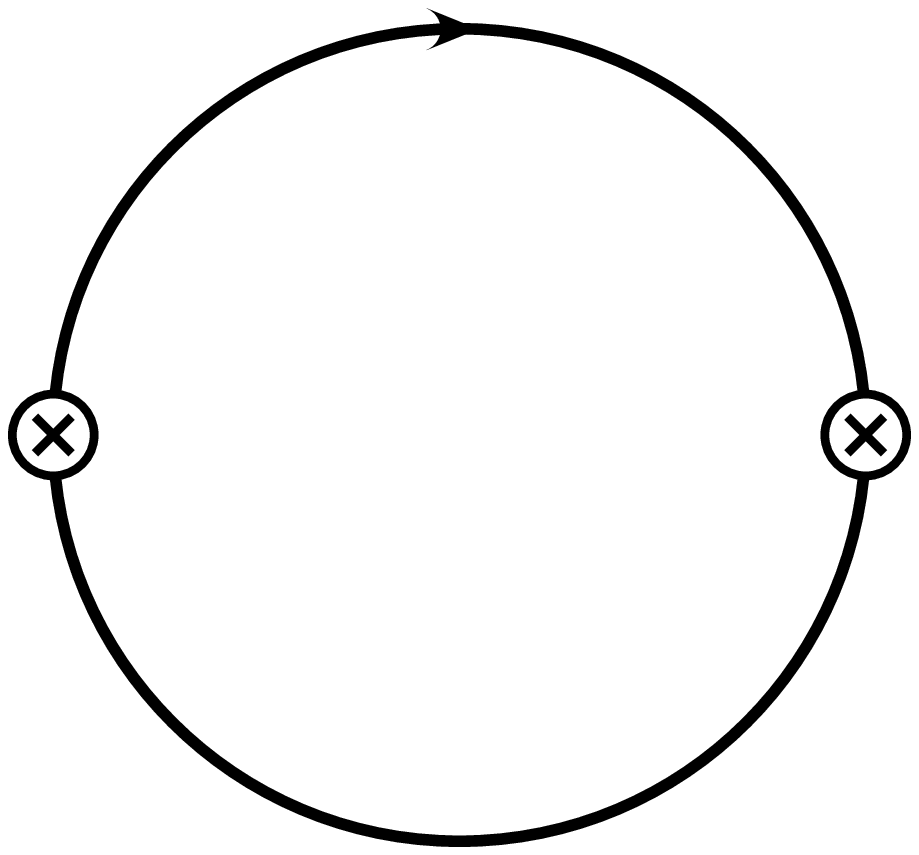}
   &
   \epsfxsize=2.5cm
   \leavevmode
   \epsffile[170 280 430 520]{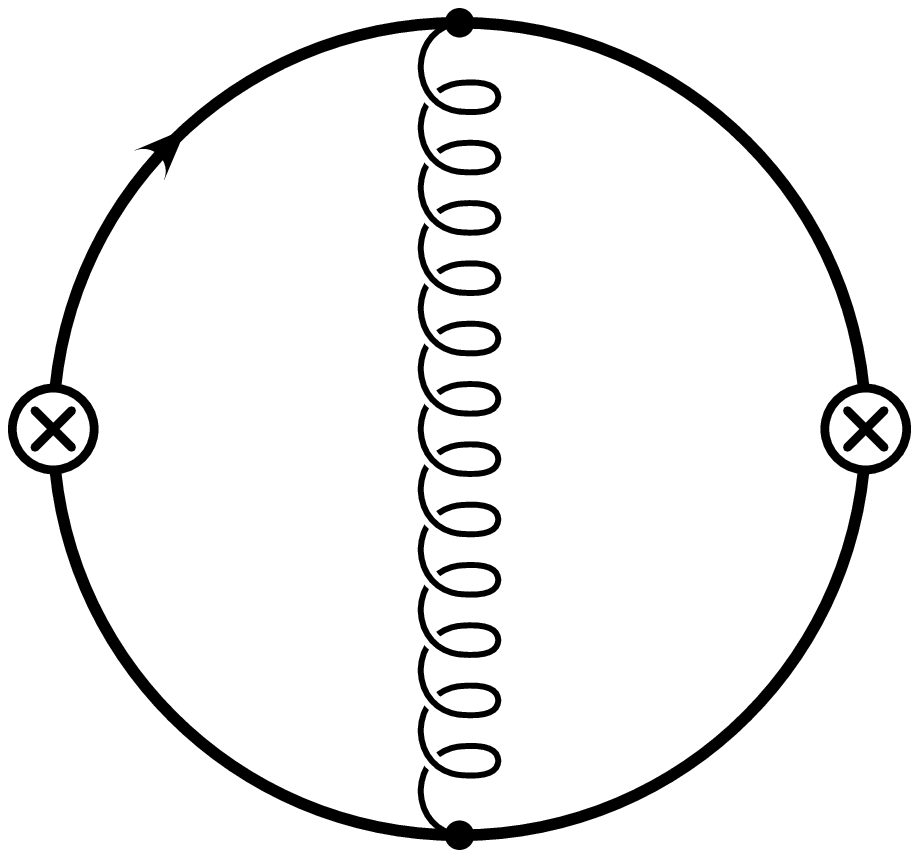}
   &
   \epsfxsize=2.5cm
   \leavevmode
   \epsffile[170 280 430 520]{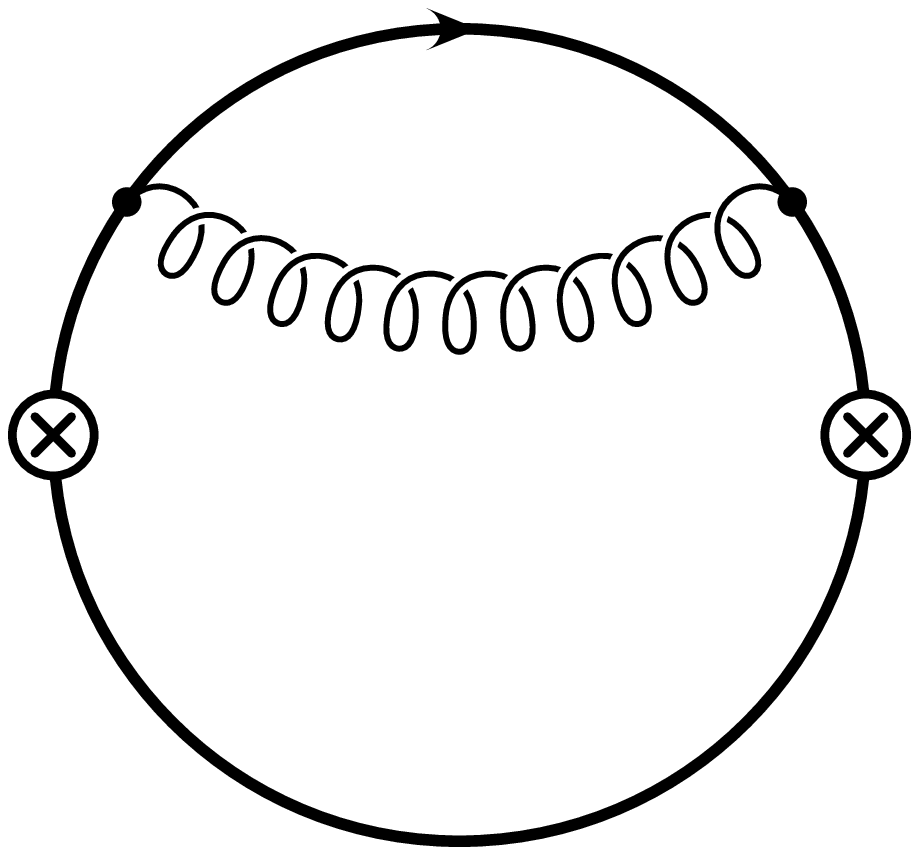}
 \end{tabular}
 \caption{\label{fig2loop} One- and two-loop diagrams 
              contributing to $\Pi^{(0)}$ and $\Pi^{(1)}$, respectively.}
 \end{center}
\end{figure}

In this paper we are only concerned with contributions to 
$\Pi(q^2)$ and $R(s)$ which originate from diagrams where
the electromagnetic current couples to the massive quark.
Diagrams where the electromagnetic current couples to a massless
quark and the massive quark is produced through a virtual
gluon have been calculated in 
\cite{HoaJezKueTeu94} and will not be discussed here. 
In order $\alpha_s$ and $\alpha_s^2$ all these amplitudes
are proportional to $Q_f^2$, the square of the charge
of the massive quark. The contribution from the diagrams,
depicted in Figure~\ref{fig2loop}, are of order $\alpha_s$, 
proportional to $C_F$ and have been calculated 
analytically \cite{KaeSab55,Kni90,BroFleTar93,BarRem73}:
\begin{eqnarray}
\Pi^{(0)} &=& \frac{3}{16\pi^2}\Bigg(
              \frac{20}{9} + \frac{4}{3z} 
           -  \frac{4(1-z)(1+2z)}{3z} G(z)
                               \Bigg),
\nonumber\\
\Pi^{(1)} &=& \frac{3}{16\pi^2}\Bigg(
              \frac{5}{6} + \frac{13}{6z}
              -\frac{(1-z)(3+2z)}{z} G(z)
              +\frac{(1-z)(1-16z)}{6z} G^2(z)
\nonumber\\
&&
              -\frac{(1+2z)}{6z}
                    \left(1+2z(1-z)\frac{d}{dz}\right)\frac{I(z)}{z}
                               \Bigg),
\end{eqnarray}
with
\begin{eqnarray}
I(z) &=& 6 \left( \zeta(3) + 4\mbox{Li}_3(u) + 2\mbox{Li}_3(u) \right)
        -8 \left( 2\mbox{Li}_2(-u) + \mbox{Li}_2(u) \right) \ln u
\nonumber\\
&&
        -2 \left( 2\ln(1+u) + \ln(1-u) \right) \ln^2 y,
\nonumber\\
&& G(z)=\frac{2u\ln u}{u^2-1},\,\,\,\,  
   u=\frac{\sqrt{1-1/z}-1}{\sqrt{1-1/z}+1},\,\,\,\,
   z=\frac{q^2}{4m^2}.
\nonumber
\end{eqnarray}

\begin{figure}[ht]
 \begin{center}
 \begin{tabular}{cc}
   \epsfxsize=2.5cm
   \leavevmode
   \epsffile[170 280 430 520]{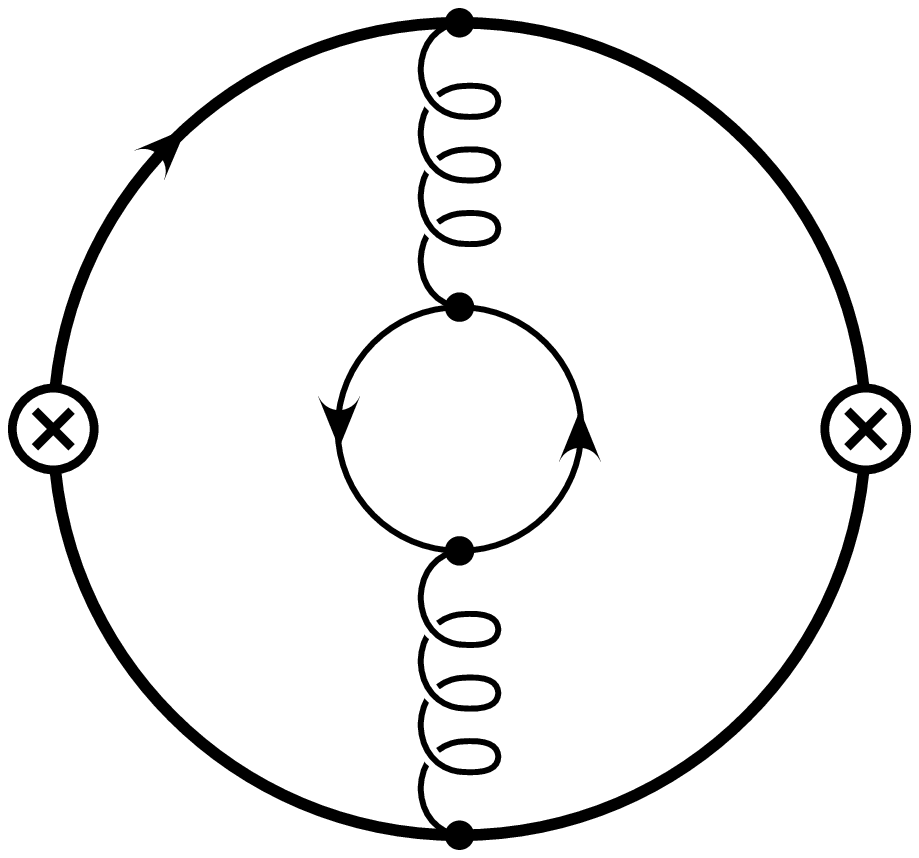}
   &
   \epsfxsize=2.5cm
   \leavevmode
   \epsffile[170 280 430 520]{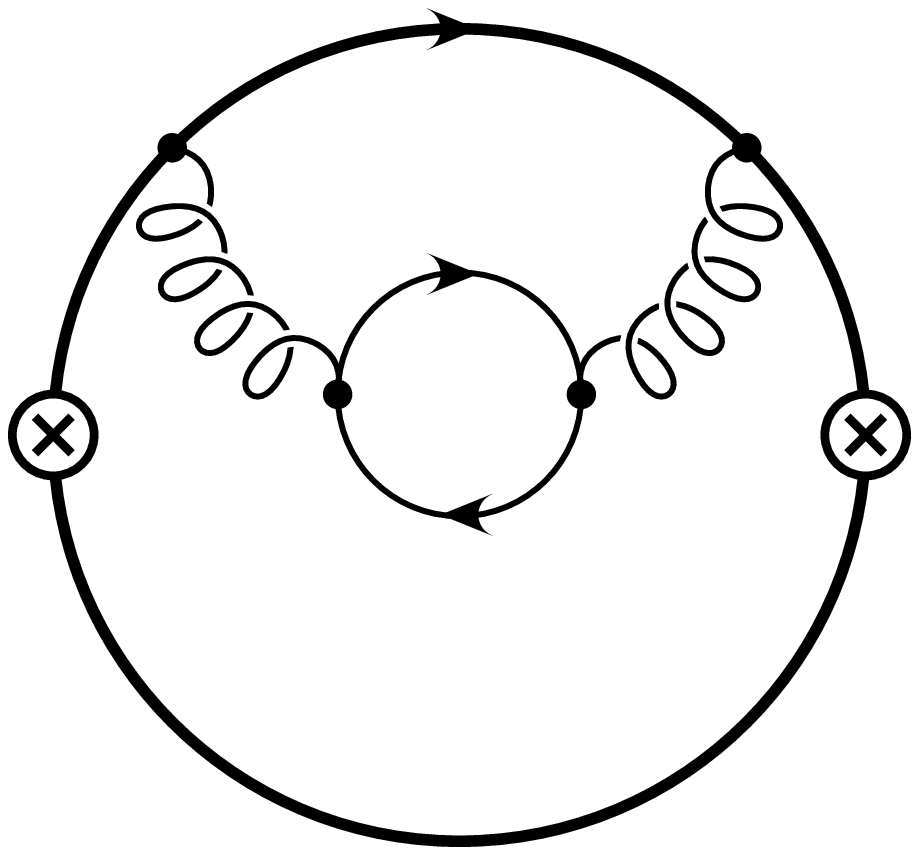}
 \end{tabular}
 \caption{\label{fig3loopnf} 
             Three-loop diagrams contributing to $\Pi^{(2)}_l$ (inner
             quark massless) and $\Pi_F^{(2)}$ (both quarks have the
             same mass $m$).}
 \end{center}
\end{figure}

\begin{figure}[ht]
 \begin{center}
 \begin{tabular}{cccc}
   \epsfxsize=2.5cm
   \leavevmode
   \epsffile[170 280 430 520]{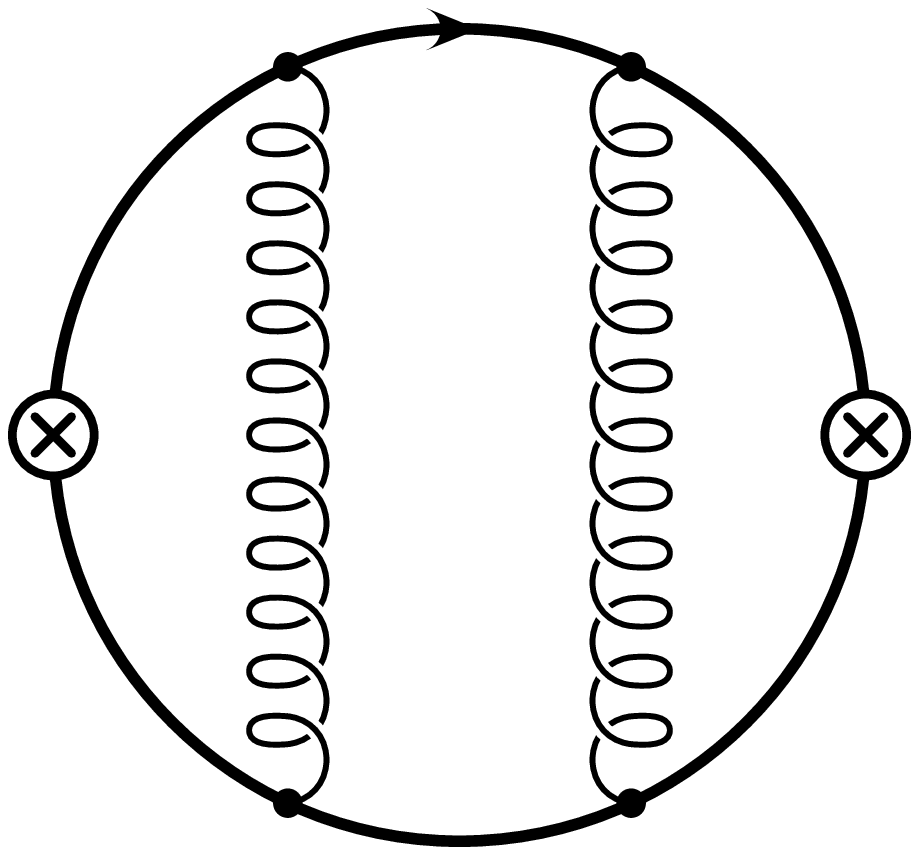}
   &
   \epsfxsize=2.5cm
   \leavevmode
   \epsffile[170 280 430 520]{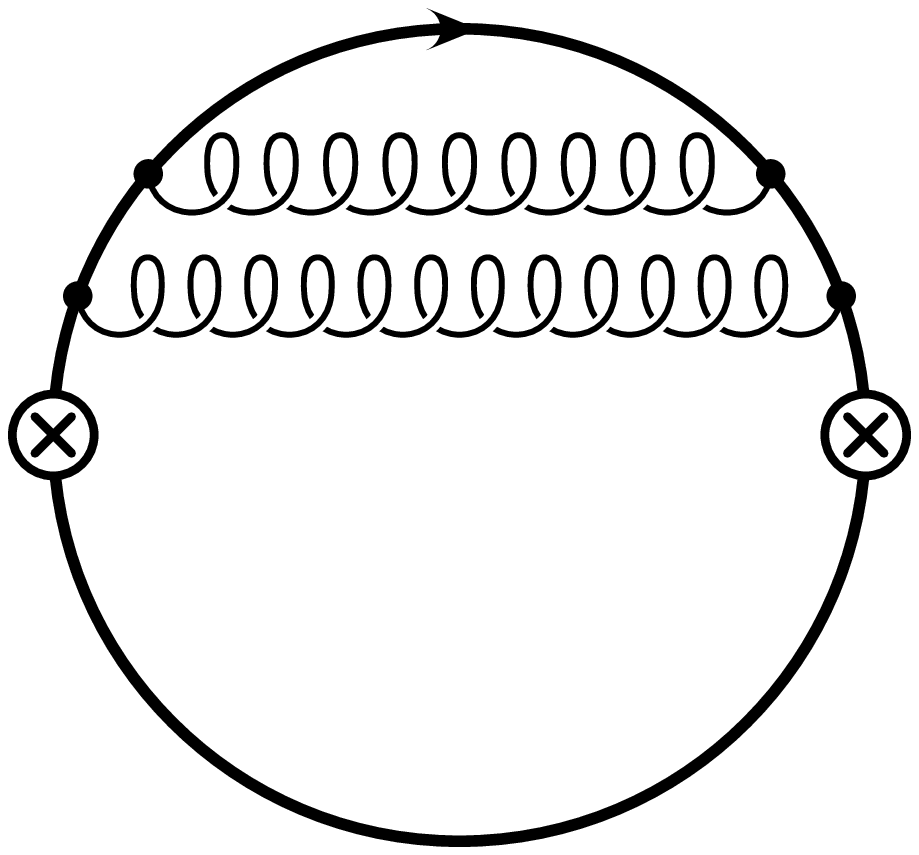}
   &
   \epsfxsize=2.5cm
   \leavevmode
   \epsffile[170 280 430 520]{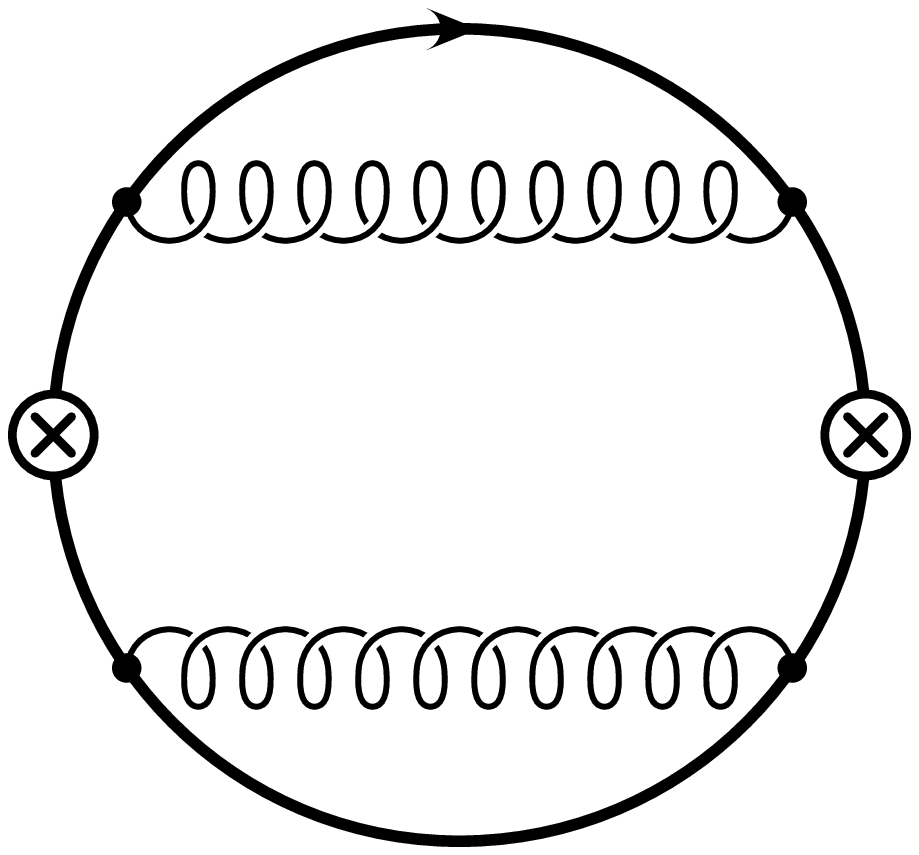}
   &
   \epsfxsize=2.5cm
   \leavevmode
   \epsffile[170 280 430 520]{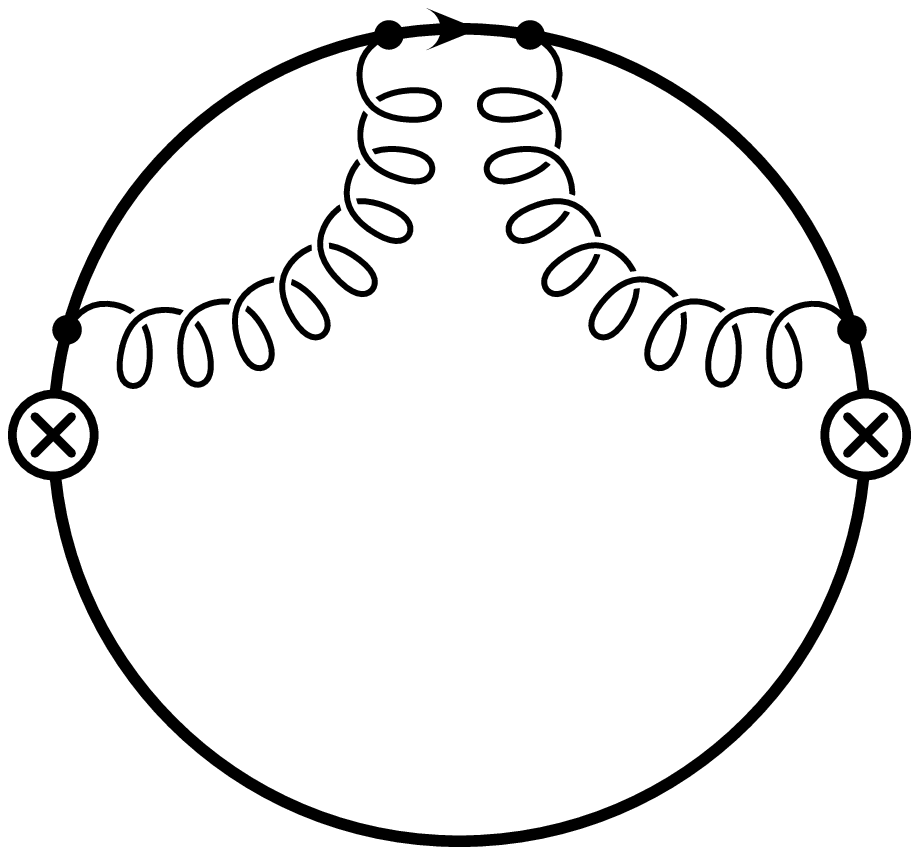}
   \\
   \epsfxsize=2.5cm
   \leavevmode
   \epsffile[170 280 430 520]{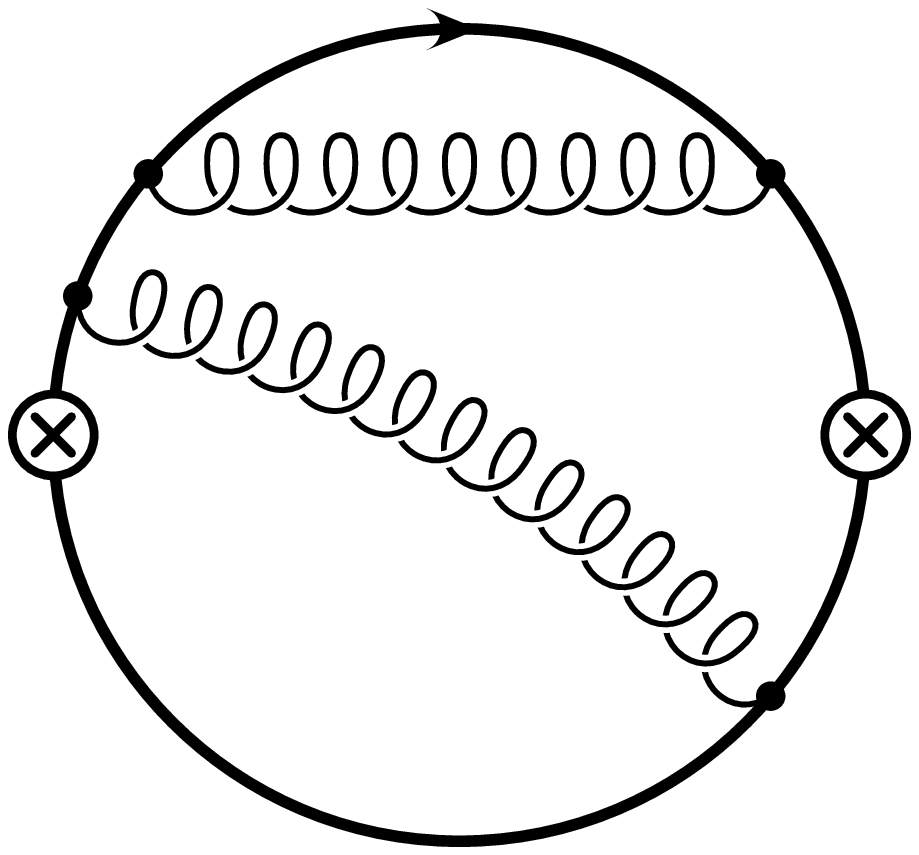}
   &
   \epsfxsize=2.5cm
   \leavevmode
   \epsffile[170 280 430 520]{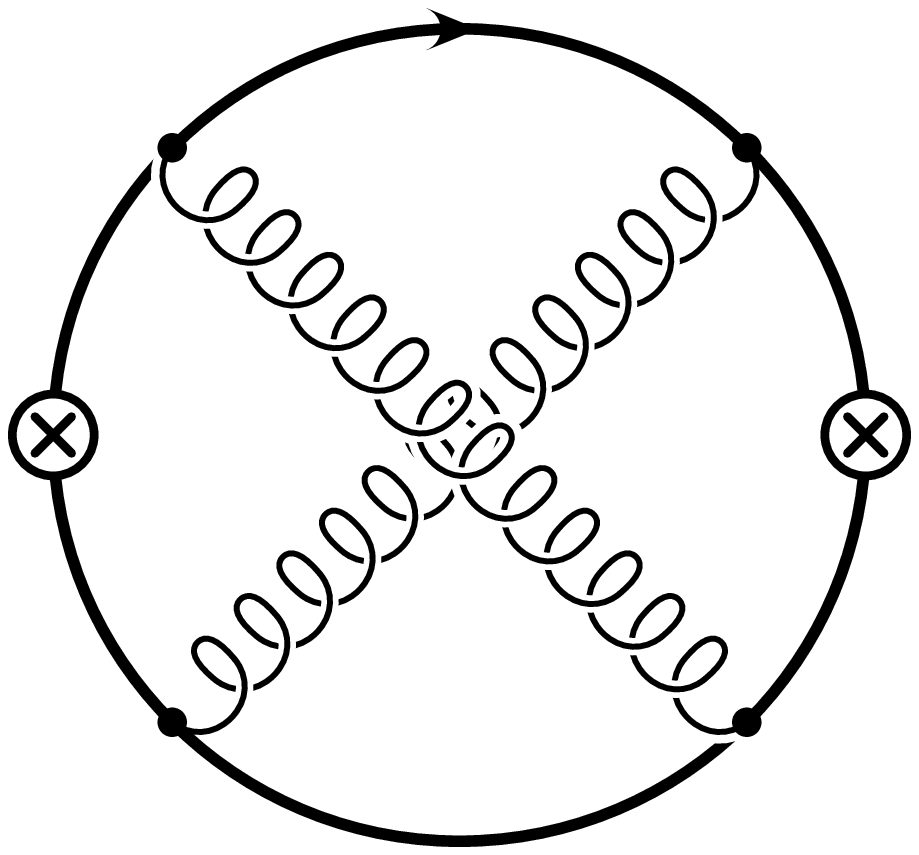}
   &
   \epsfxsize=2.5cm
   \leavevmode
   \epsffile[170 280 430 520]{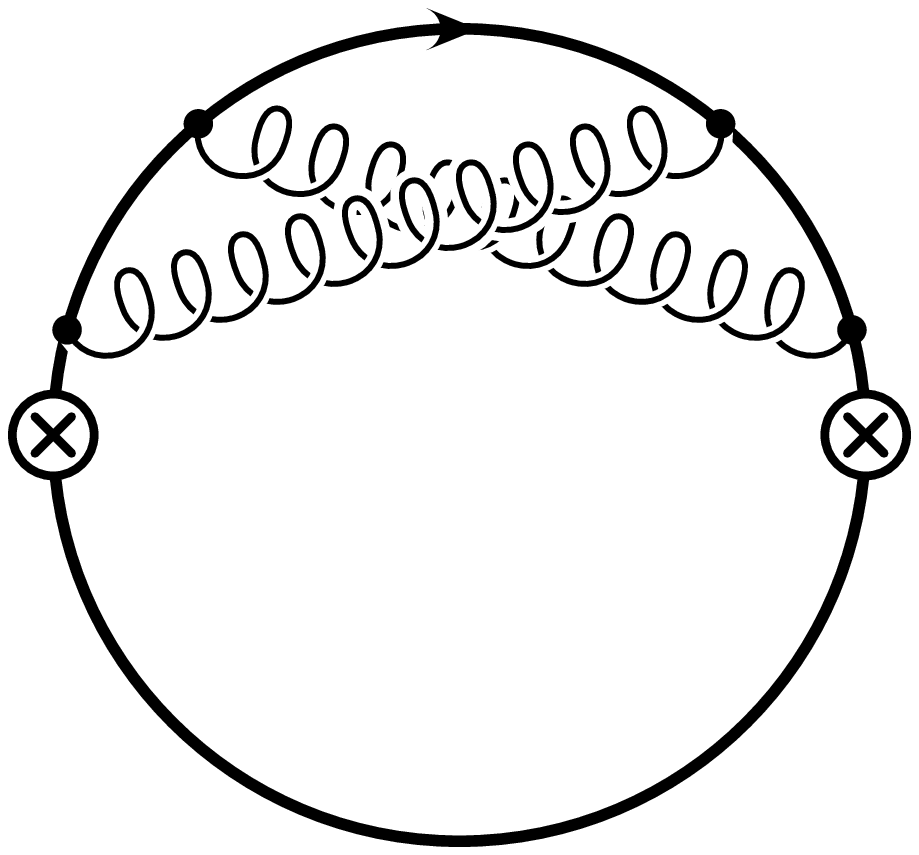}
   &
   \epsfxsize=2.5cm
   \leavevmode
   \epsffile[170 280 430 520]{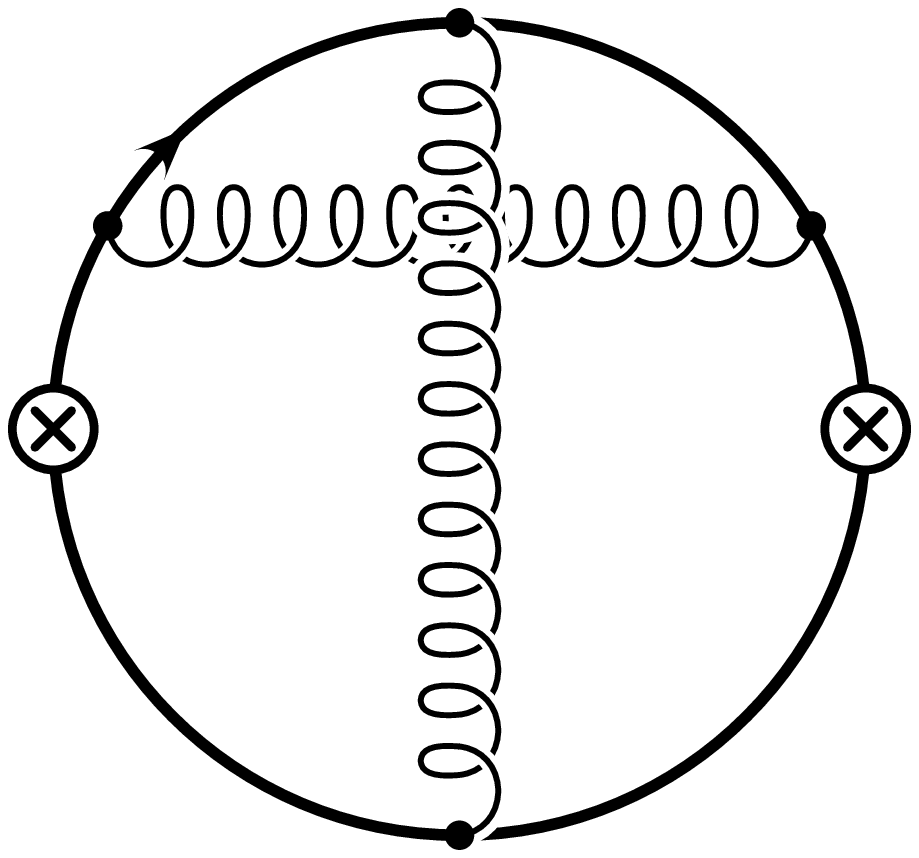}
   \\
   \epsfxsize=2.5cm
   \leavevmode
   \epsffile[170 280 430 520]{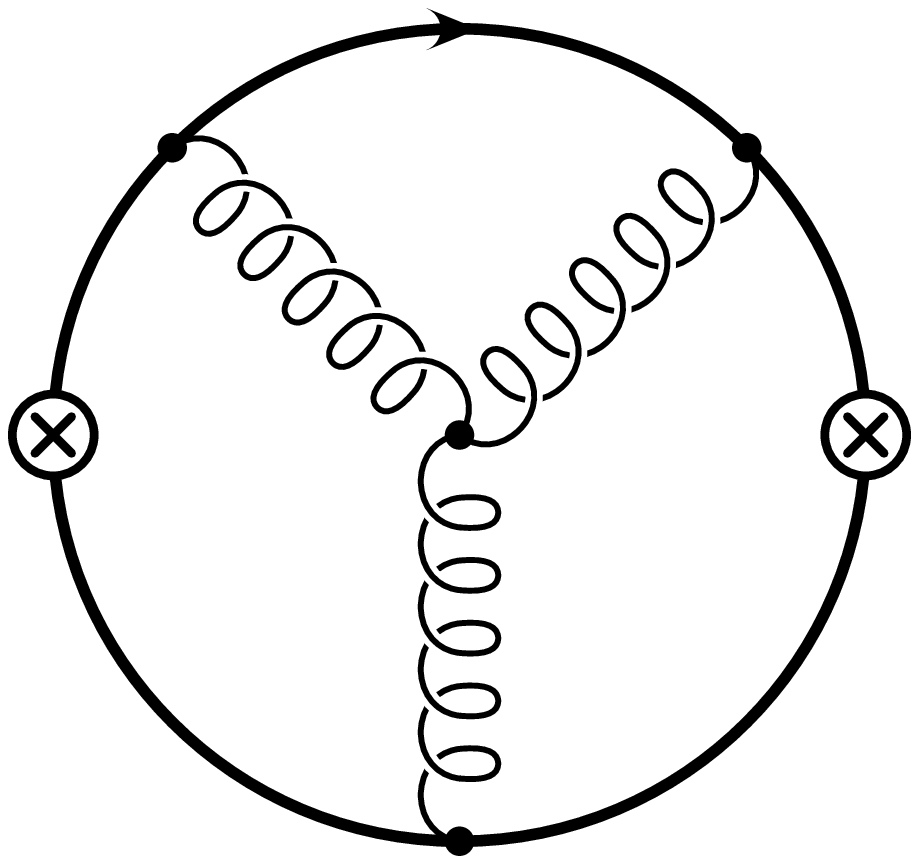}
   &
   \epsfxsize=2.5cm
   \leavevmode
   \epsffile[170 280 430 520]{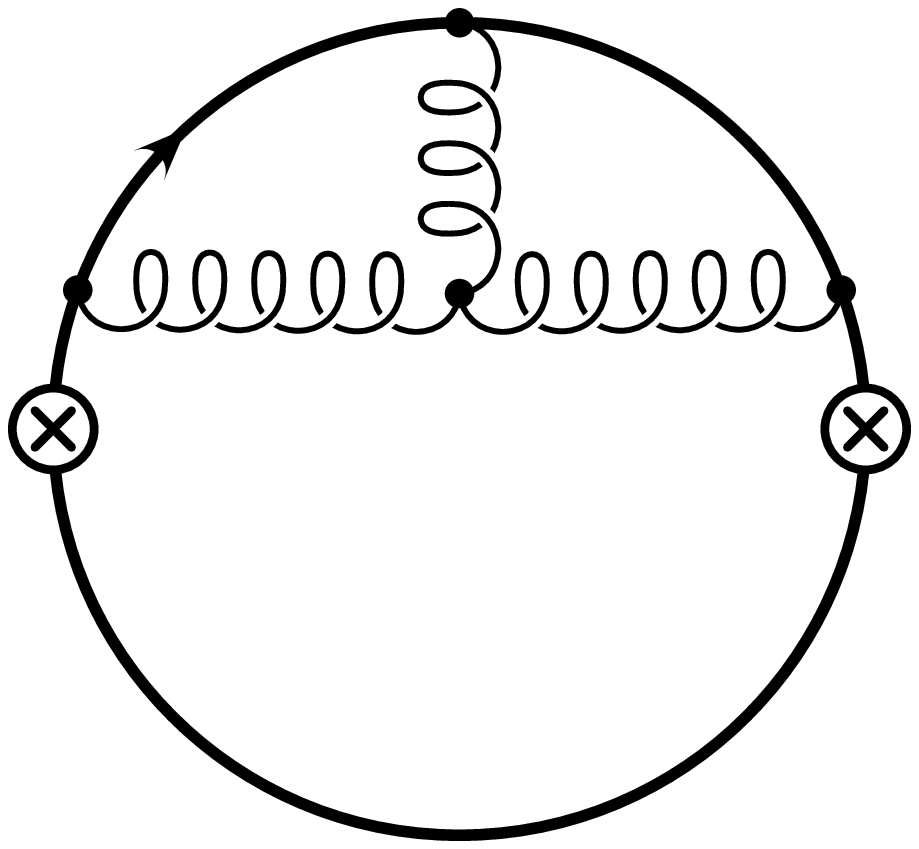}
   &
   \epsfxsize=2.5cm
   \leavevmode
   \epsffile[170 280 430 520]{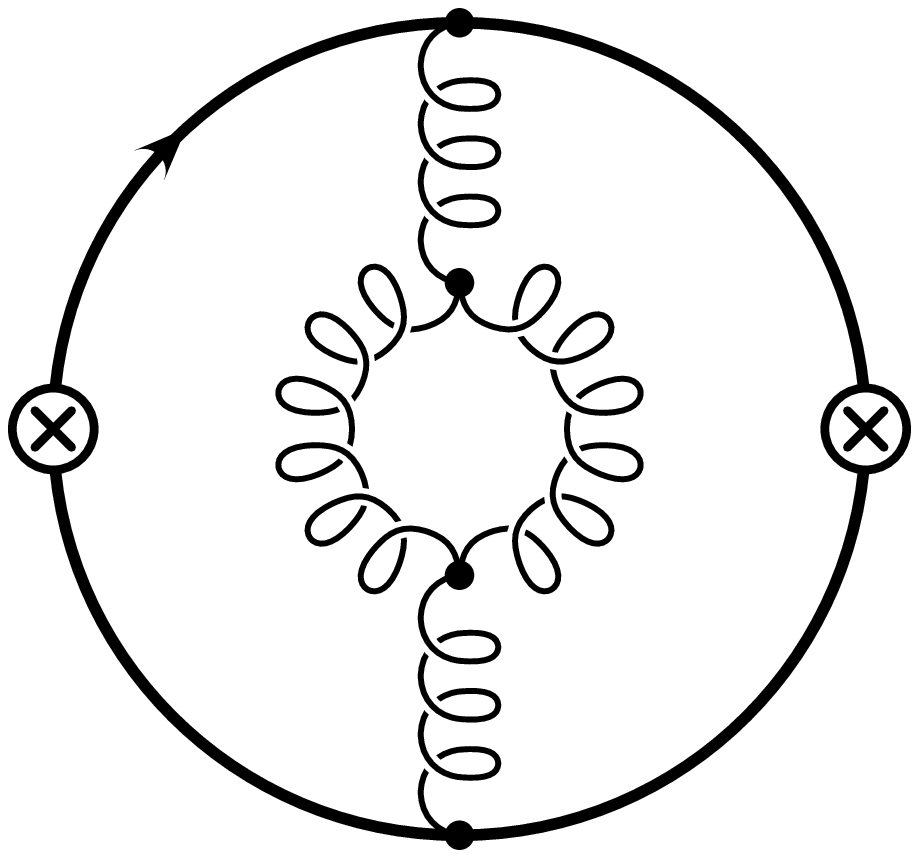}
   &
   \epsfxsize=2.5cm
   \leavevmode
   \epsffile[170 280 430 520]{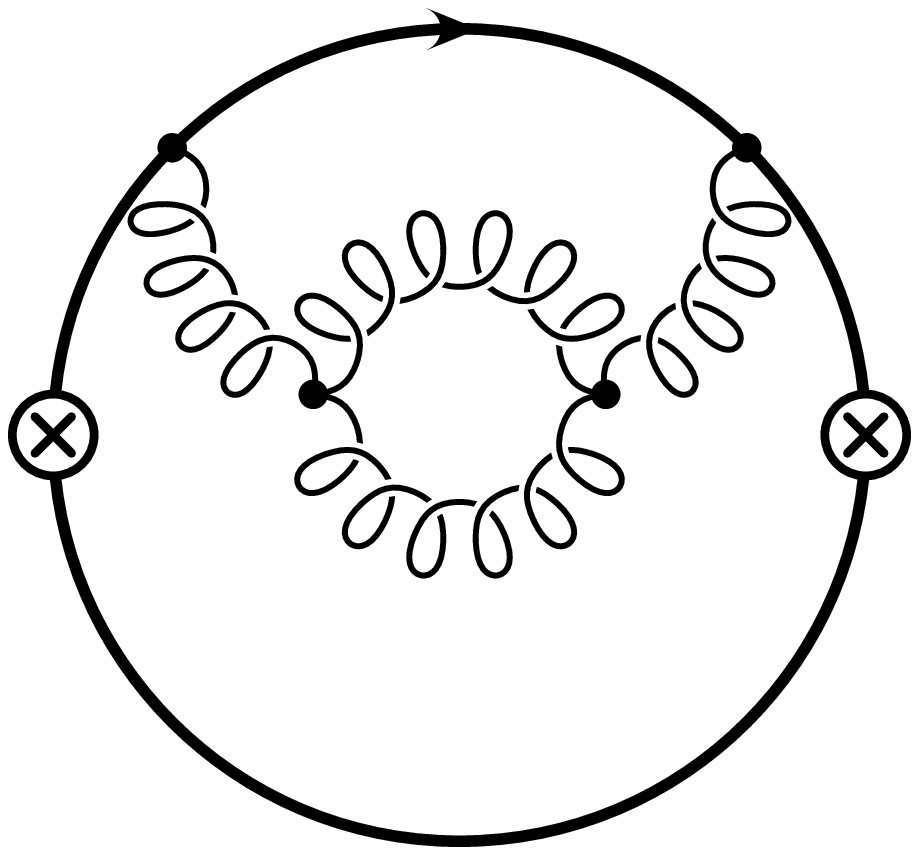}
   \\
 \end{tabular}
 \caption{\label{fig3loop} Purely gluonic contribution to 
         ${\cal O}(\alpha_s^2)$. Diagrams with ghost loops are
         not depicted.}
 \end{center}
\end{figure}

For the $\alpha_s^2$ calculation the diagrams can be 
decomposed according to the colour structure.
The contributions from diagrams with $n_l$ light 
or one massive 
internal fermion loop will be denoted
by $C_F T n_l\Pi_l$ and 
$C_F T\Pi_F$ with the group theoretical
coefficients
factored out
(Figure~\ref{fig3loopnf}). 
Purely gluonic corrections (Figure~\ref{fig3loop})
are proportional to $C_F^2$ or $C_F C_A$. The former are the only
contributions in an Abelian theory, the latter are characteristic for
the non-Abelian aspects of QCD. It will be important in the 
subsequent discussion to 
treat these two classes separately, since they exhibit strikingly 
different behaviour close to threshold. The following decomposition
of $\Pi(q^2)$ is therefore adopted throughout the paper
\begin{eqnarray}
\Pi &=& \Pi^{(0)} + \frac{\alpha_s(\mu^2)}{\pi} C_F \Pi^{(1)} 
\nonumber\\
&&
       + \left(\frac{\alpha_s(\mu^2)}{\pi}\right)^2
         \left[ C_F^2       \Pi_A^{(2)}
              + C_A C_F     \Pi_{\it NA}^{(2)}
              + C_F T   n_l \Pi_l^{(2)}
              + C_F T       \Pi_F^{(2)}
         \right].
\end{eqnarray}
All steps described below will be performed separately for
the first three $\Pi^{(2)}$. In fact, new information will only be 
obtained for $\Pi_A^{(2)}$ 
and $\Pi_{\it NA}^{(2)}$.
The imaginary part of
$\mbox{Im}\Pi_l^{(2)}$ 
is known analytically already
\cite{HoaKueTeu95}.
$\mbox{Im}\Pi_F^{(2)}$ can easily be calculated numerically
for the following reasons:
The amplitude with a massive internal fermion exhibits a two
particle cut with threshold at $2m$ which can be calculated analytically
\cite{HoaKueTeu95}
and a four particle cut with threshold at $4m$ which is given
in terms of a two dimensional integral
\cite{HoaKueTeu95}
which can be solved 
easily numerically.
This contribution is therefore known already and 
will not be treated in this paper.

For convenience of the reader we give in the remaining part 
of this introduction a short description of the individual steps 
of the calculation.

A crucial input for our procedure is the behaviour of
$\Pi(q^2)$ for $q^2\to -\infty$. In Section \ref{seclar}
we list the terms of order $(m^2/q^2)^0$ and $(m^2/q^2)^1$
for $\Pi^{(2)}$ in the $\overline{\mbox{MS}}$
and on-shell scheme. A very important test for the
correctness of the result is
provided by the $m^4/s^2$ terms which are only known for $R(s)$,
i.e. the imaginary part of $\Pi$.

In Section \ref{secthr} we discuss the behaviour of $\Pi(q^2)$ at
threshold. For the QED-like corrections 
the leading $1/v$ singularities are
resummed via Sommerfeld's formula: $R^v=3vx_s/2(1-e^{-x_s})$
with $x_s=C_F\pi\alpha_s/v$. An expansion up to 
${\cal O}(\alpha_s^2)$ delivers the leading singularity
for $\Pi_A^{(2)}(q^2\to 4m^2)$ and in combination with the
``hard'' one-loop vertex correction also the subleading
logarithmic contribution.
With the interpretation of $\alpha_s$ in $R^v$ as $\alpha_V(4\vec{p}\,^2)$,
with $\vec{p}$ being is the momentum of one quark,
Eq.~(\ref{alphav}) reproduces on one side the known threshold
term for $R_l^{(2)}$ and leads on the other side to an expression
for the leading logarithm of $R_{\it NA}^{(2)}$. 

Important information is contained in the Taylor series of
$\Pi(q^2)$ around zero. The first seven nontrivial terms are 
computed. The calculation is based on the evaluation of 
three-loop tadpole integrals which is performed with the
program package MATAD written in FORM
\cite{VerFORM}. In Section \ref{secsma} we describe the method
in more detail and present our results.

Section \ref{secQED}
is devoted to a discussion of the scheme 
dependence of the polarization function as well as to a related question of
the connection between the QED coupling constant 
$\alpha$ in the on-shell and
the $\overline{\mbox{MS}}$ scheme. 
We explicitly derive the connection up to and including the
terms of order $\alpha^2 \alpha_s^2$ using the results of the
previous section. 

Section \ref{secoptappr} contains a description of the
procedure which combines the results from the different kinematical
regions. In a first step  a function with a nonsingular behaviour
at high energies is constructed from the known information
about $\Pi(q^2)$. Also the logarithms at
threshold are subtracted. The remaining $1/v$ singularity for
$\Pi_A^{(2)}$ will be removed by multiplication with $v$. 
The conformal mapping $z=4\omega/(1+\omega)^2$ of the complex
$q^2$-plane into the interior of the unit circle provides
a good starting point for the Pad\'e approximation method.
The different Pad\'e approximants are denoted by $[n/m]$
and are rational functions with a polynomial of degree
$n$ in the numerator and $m$ in the denominator.

The results are discussed and explicit
formulae and handy approximations for $R_A$ and $R_{\it NA}$ are given
in Section \ref{secres}.
Section \ref{seccon} finally contains our conclusions.

%%%%%%%%%%%%%%%%%%%%%%%%%%%%%%%%%%%%%%%%%%%%%%%%%%%%%%%%%%%%
%%%%%%%%%%%%%%%%%%%%%%%%%%%%%%%%%%%%%%%%%%%%%%%%%%%%%%%%%%%%

\mathversion{bold}
\section{\label{seclar}Large $q^2$ Behaviour}
\mathversion{normal}

The expansion of $\Pi(q^2)$ for large $q^2$ provides an important
constraint on $\Pi$ in the entire complex plane. For
the imaginary part it has been argued
\cite{CheKue94}
that the inclusion of the constant, plus quadratic 
plus quartic terms in $m$
provides an adequate approximation down to fairly close to
threshold. 
The large $q^2$ expansion of $\Pi(q^2)$ reads as follows
(The terms proportional to $C_F T$ are listed for completeness.):
\begin{eqnarray}
\Pi^{(0)} &=&  \frac{3}{16\pi^2}\Bigg[
          \frac{20}{9} -\frac{4}{3}\ln\frac{-q^2}{m^2}
         +8\frac{m^2}{q^2}
         +\left(\frac{4m^2}{q^2}\right)^2
          \left(\frac{1}{4}+\frac{1}{2}\ln\frac{-q^2}{m^2}\right)
         + \ldots
                                \Bigg],
\label{pi1l}
\\
\Pi^{(1)} &=&  \frac{3}{16\pi^2}\Bigg[
          \frac{5}{6} - 4\zeta(3) - \ln\frac{-q^2}{m^2} 
         -12\frac{m^2}{q^2}\ln\frac{-q^2}{m^2} 
\nonumber\\
&&
         +\left(\frac{4m^2}{q^2}\right)^2
          \left(\frac{1}{24} + \zeta(3) - \frac{5}{8}\ln\frac{-q^2}{m^2} 
                                        - \frac{3}{4}\ln^2\frac{-q^2}{m^2} 
          \right)
         + \ldots
                                \Bigg],
\label{pi2l}
\\
\Pi_A^{(2)}&=&\frac{3}{16\pi^2}\Bigg[
               -\frac{121}{48} -5\zeta(2) + 8\zeta(2)\ln2 
               -\frac{99}{16}\zeta(3) + 10\zeta(5) 
               +\frac{1}{8}\ln\frac{-q^2}{m^2} 
\nonumber\\
&&
+\frac{4m^2}{q^2}\left(
                \frac{139}{12}
               -\frac{15}{2}\zeta(2)
               +12\zeta(2)\ln2
               -\frac{41}{12}\zeta(3)
               -\frac{35}{6}\zeta(5)
\right.
\nonumber\\
&&
\left.
               -\frac{3}{8}\ln\frac{-q^2}{m^2} 
               +\frac{9}{4}\ln^2\frac{-q^2}{m^2} 
\right)                                
\Bigg]
+ \ldots,
\label{pim2a}
\\
\Pi_{\it NA}^{(2)}&=&\frac{3}{16\pi^2}\Bigg[
        \frac{4873}{432} + \frac{4}{3}\zeta(2) - 4\zeta(2)\ln2 
       -\frac{4013}{288}\zeta(3)  - \frac{5}{3}\zeta(5) 
       - \frac{157}{36}\ln\frac{-q^2}{m^2} 
\nonumber\\
&&
       + \frac{11}{12}\ln\frac{-q^2}{\mu^2}\ln\frac{-q^2}{m^2}  
       - \frac{11}{24}\ln^2\frac{-q^2}{m^2}  
       - \frac{55}{72}\ln\frac{-q^2}{\mu^2} 
       + \frac{11}{3}\zeta(3)\ln\frac{-q^2}{\mu^2} 
\nonumber\\
&&
+\frac{4m^2}{q^2}\left(
         \frac{7}{2} + 2\zeta(2) - 6\zeta(2)\ln2 
       + \frac{17}{6}\zeta(3) - \frac{85}{12}\zeta(5)
\right.
\nonumber\\
&&
\left.
       - \frac{185}{24}\ln\frac{-q^2}{m^2} 
       + \frac{11}{4}\ln\frac{-q^2}{\mu^2}\ln\frac{-q^2}{m^2}  
       - \frac{11}{8}\ln^2\frac{-q^2}{m^2}  
\right)
\Bigg]
+ \ldots,
\label{pim2na}
\\
\Pi_l^{(2)}&=&\frac{3}{16\pi^2}\Bigg[
    - \frac{116}{27}  
    + \frac{4}{3}\zeta(2) 
    + \frac{38}{9}\zeta(3) 
    + \frac{14}{9}\ln\frac{-q^2}{m^2}
    - \frac{1}{3}\ln\frac{-q^2}{\mu^2}\ln\frac{-q^2}{m^2}
\nonumber\\
&&
    + \frac{1}{6}\ln^2\frac{-q^2}{m^2}
    + \frac{5}{18}\ln\frac{-q^2}{\mu^2}
    - \frac{4}{3}\zeta(3)\ln\frac{-q^2}{\mu^2}
\nonumber\\
&&
+\frac{4m^2}{q^2}\left(
    -1 + 2\zeta(2) 
    + \frac{13}{6}\ln\frac{-q^2}{m^2}
    - \ln\frac{-q^2}{\mu^2}\ln\frac{-q^2}{m^2}
    + \frac{1}{2}\ln^2\frac{-q^2}{m^2}
\right)
\Bigg]
+ \ldots,
\label{pim2f}
\\
\Pi_F^{(2)}&=&
\frac{3}{16\pi^2}\Bigg[
    - \frac{307}{216}  
    - \frac{8}{3}\zeta(2) 
    + \frac{545}{144}\zeta(3) 
    + \frac{14}{9}\ln\frac{-q^2}{m^2}
    - \frac{1}{3}\ln\frac{-q^2}{\mu^2}\ln\frac{-q^2}{m^2}
\nonumber\\
&&
    + \frac{1}{6}\ln^2\frac{-q^2}{m^2}
    + \frac{5}{18}\ln\frac{-q^2}{\mu^2}
    - \frac{4}{3}\zeta(3)\ln\frac{-q^2}{\mu^2}
+\frac{4m^2}{q^2}\left(
    - \frac{10}{3} 
    - 4\zeta(2)
\right.
\nonumber\\
&&
\left.
    + 4\zeta(3) 
    + \frac{13}{6}\ln\frac{-q^2}{m^2}
    - \ln\frac{-q^2}{\mu^2}\ln\frac{-q^2}{m^2}
    + \frac{1}{2}\ln^2\frac{-q^2}{m^2}
\right)
\Bigg]
+ \ldots.
\label{pim2f2}
\end{eqnarray}
(The logarithmic contributions of the $(m^2/q^2)^0$ terms
were originally obtained in 
\cite{CheKatTka79DinSap79CelGon80}, 
while the constants in $\Pi^{(2)}$ 
were first published in 
\cite{GorKatLar91,SurSam91}, albeit not in
an explicit form\footnote{For instance, in \cite{GorKatLar91} 
        the relation following Eq.~(8) expresses  
        the constant contribution of $\Pi^{(2)}$ by some 
        quantities which are explicitly given in this article.}.
The quadratic mass terms were first found in 
\cite{GorKatLar86} and later confirmed in
\cite{CheKwi93}.)
Eqs.~(\ref{pim2a}-\ref{pim2f}) imply the
following contributions to $R$:
\begin{eqnarray}
R^{(0)} &=&  3\left[1 - 6 \frac{m^4}{s^2}+ \ldots\right],
\\
R^{(1)} &=&  3\left[ \frac{3}{4}
                   + 9 \frac{m^2}{s}           
                   + \frac{m^4}{s^2}\left(\frac{15}{2}
                                         -18\ln\frac{m^2}{s}\right)
                   + \ldots
              \right],
\\
R_A^{(2)}&=& 3 \left[
           -\frac{3}{32}
           +\frac{m^2}{s}
            \left(\frac{9}{8}
                 +\frac{27}{2}\ln\frac{m^2}{s}
            \right)
                + \ldots \right],
\label{rm2a}
\\
R_{\it NA}^{(2)}&=& 3 \left[
            \frac{123}{32}
           -\frac{11}{4}\zeta(3)
           +\frac{11}{16}\ln\frac{\mu^2}{s}
           +\frac{m^2}{s}
            \left(\frac{185}{8}
                 +\frac{33}{4}\ln\frac{\mu^2}{s}
            \right)
                + \ldots \right],
\\
R_l^{(2)}&=& 3 \left[
           -\frac{11}{8}
           +\zeta(3)
           -\frac{1}{4}\ln\frac{\mu^2}{s}
           +\frac{m^2}{s}
            \left(-\frac{13}{2}
                 -3\ln\frac{\mu^2}{s}
            \right)
                + \ldots \right].
\label{rm2l}
\end{eqnarray}
Note that $R^{(2)}_F\equiv R^{(2)}_l $ if one considers  only
$m^0$ and $m^2$ mass corrections. The equality can be understood
within the renormalization group approach of 
\cite{CheKue90}.

The quartic contributions to $R$ have been obtained in
\cite{CheKue94}.
They will be important for comparisons with the present results and
read as follows:
\begin{eqnarray}
R_A^{(2)\scriptsize, m^4}&=&
3\frac{m^4}{s^2}\Bigg[
      -\frac{345}{16} + 99\zeta(2) - 72\ln2\zeta(2) + 36\zeta(3)
\nonumber\\
&&
      +\frac{81}{4}\ln\frac{m^2}{s}
      -27\ln^2\frac{m^2}{s}
              \Bigg],
\label{rm4a}
\\
R_{\it NA}^{(2)\scriptsize, m^4}&=&
3\frac{m^4}{s^2}\Bigg[
       \frac{77}{3} + \frac{9}{2}\zeta(2) + 36\ln2\zeta(2) + 11\zeta(3)
\nonumber\\
&&
      -\frac{381}{8}\ln\frac{m^2}{s}
      +\frac{33}{4}\ln^2\frac{m^2}{s}
      +\frac{55}{8}\ln\frac{\mu^2}{s}
      -\frac{33}{2}\ln\frac{m^2}{s}\ln\frac{\mu^2}{s}
              \Bigg],
\\
R_l^{(2)\scriptsize, m^4}&=&
3\frac{m^4}{s^2}\Bigg[
      -\frac{35}{6} - 18\zeta(2) - 4\zeta(3)
\nonumber\\
&&
      +\frac{27}{2}\ln\frac{m^2}{s}
      -3\ln^2\frac{m^2}{s}
      -\frac{5}{2}\ln\frac{\mu^2}{s}
      +6\ln\frac{m^2}{s}\ln\frac{\mu^2}{s}
              \Bigg],
\label{rm4l}\\
R_F^{(2)\scriptsize, m^4}&=&
3\frac{m^4}{s^2}\Bigg[
       \frac{2}{3} + 18\zeta(2) - 10\zeta(3)
\nonumber\\
&&
      +12\ln\frac{m^2}{s}
      -3\ln^2\frac{m^2}{s}
      -\frac{5}{2}\ln\frac{\mu^2}{s}
      +6\ln\frac{m^2}{s}\ln\frac{\mu^2}{s}
              \Bigg].
\end{eqnarray}

The results have been expressed in terms of the on-shell mass in order to 
combine them easily with the threshold and low $q^2$ behaviour. 
As can be seen from Eqs.~(\ref{pim2a}), (\ref{rm2a}) 
and (\ref{rm4a}) the 
Abelian parts of $\Pi^{(2)}$ and $R^{(2)}$ are
in this case independent of $\mu$ whereas the 
$\mu$ dependence of non-Abelian and fermionic contributions 
is connected to the
running of the coupling constant $\alpha_s(\mu^2)$.
For completeness the 
Eqs.~(\ref{pi1l}-\ref{pim2f2}) are reexpressed 
in terms of the $\overline{\mbox{MS}}$ mass $\bar{m}$:
\begin{eqnarray}
\bar{\Pi}^{(0)} &=&  \frac{3}{16\pi^2}\Bigg[
          \frac{20}{9} -\frac{4}{3}\ln\frac{-q^2}{\mu^2}
         +8\frac{\bar{m}^2}{q^2}
         +\left(\frac{4\bar{m}^2}{q^2}\right)^2
          \left(\frac{1}{4}+\frac{1}{2}\ln\frac{-q^2}{\bar{m}^2}\right)
         + \ldots
                                \Bigg],
\\
\bar{\Pi}^{(1)}&=&\frac{3}{16\pi^2}\Bigg[
     \frac{55}{12} - 4\zeta(3)  
   - \ln\frac{-q^2}{\mu^2}
   +\frac{4\bar{m}^2}{q^2}\left(
          4
        - 3\ln\frac{-q^2}{\mu^2}
                    \right)
\nonumber\\
&&
+\left(\frac{4\bar{m}^2}{q^2}\right)^2\left(
       \frac{1}{24}
      +\zeta(3)
      +\frac{11}{8}\ln\frac{-q^2}{\bar{m}^2}
      +\frac{3}{4}\ln^2\frac{-q^2}{\bar{m}^2}
      -\frac{3}{2}\ln\frac{-q^2}{\bar{m}^2}\ln\frac{-q^2}{\mu^2}
\right)   
\Bigg],
\\
\bar{\Pi}_A^{(2)}&=&\frac{3}{16\pi^2}\Bigg[
       - \frac{143}{72} 
       - \frac{37}{6}\zeta(3)
       + 10\zeta(5)
       + \frac{1}{8}\ln\frac{-q^2}{\mu^2}
\nonumber\\
&&
+\frac{4\bar{m}^2}{q^2}\left(
       \frac{1667}{96} 
     - \frac{5}{12}\zeta(3)
     - \frac{35}{6}\zeta(5)
     - \frac{51}{8}\ln\frac{-q^2}{\mu^2}
     + \frac{9}{4}\ln^2\frac{-q^2}{\mu^2}
\right)                                
\Bigg]
+ \ldots,
\label{pim2ams}
\\
\bar{\Pi}_{\it NA}^{(2)}&=&\frac{3}{16\pi^2}\Bigg[
       + \frac{44215}{2592} 
       - \frac{227}{18}\zeta(3)
       - \frac{5}{3}\zeta(5)
%%%\nonumber\\
%%%&&
       - \frac{41}{8}\ln\frac{-q^2}{\mu^2}
       + \frac{11}{24}\ln^2\frac{-q^2}{\mu^2}
       + \frac{11}{3}\zeta(3)\ln\frac{-q^2}{\mu^2}
\nonumber\\
&&
+\frac{4\bar{m}^2}{q^2}\left(
       + \frac{1447}{96} 
       + \frac{4}{3}\zeta(3)
       - \frac{85}{12}\zeta(5)
       - \frac{185}{24}\ln\frac{-q^2}{\mu^2}
       + \frac{11}{8}\ln^2\frac{-q^2}{\mu^2}
\right)                                
\Bigg]
+ \ldots,
\label{pim2nams}
\\
\bar{\Pi}_l^{(2)}&=&\frac{3}{16\pi^2}\Bigg[
       - \frac{3701}{648} 
       + \frac{38}{9}\zeta(3)
%%%\nonumber\\
%%%&&
       + \frac{11}{6}\ln\frac{-q^2}{\mu^2}
       - \frac{1}{6}\ln^2\frac{-q^2}{\mu^2}
       - \frac{4}{3}\zeta(3)\ln\frac{-q^2}{\mu^2}
\nonumber\\
&&
+\frac{4\bar{m}^2}{q^2}\left(
       - \frac{95}{24} 
       + \frac{13}{6}\ln\frac{-q^2}{\mu^2}
       - \frac{1}{2}\ln^2\frac{-q^2}{\mu^2}
\right)                                
\Bigg]
+ \ldots,
\label{pim2lms}
\\
\bar{\Pi}_F^{(2)}&=&
\frac{3}{16\pi^2}\Bigg[
       - \frac{3701}{648} 
       + \frac{38}{9}\zeta(3)
%%%\nonumber\\
%%%&&
       + \frac{11}{6}\ln\frac{-q^2}{\mu^2}
       - \frac{1}{6}\ln^2\frac{-q^2}{\mu^2}
       - \frac{4}{3}\zeta(3)\ln\frac{-q^2}{\mu^2}
\nonumber\\
&&
+\frac{4\bar{m}^2}{q^2}\left(
       - \frac{223}{24} 
       + 4\zeta(3)
       + \frac{13}{6}\ln\frac{-q^2}{\mu^2}
       - \frac{1}{2}\ln^2\frac{-q^2}{\mu^2}
\right)                                
\Bigg]
+ \ldots.
\label{pim2Fms}
\end{eqnarray}

%%%%%%%%%%%%%%%%%%%%%%%%%%%%%%%%%%%%%%%%%%%%%%%%%%%%%%%%%%%%
%%%%%%%%%%%%%%%%%%%%%%%%%%%%%%%%%%%%%%%%%%%%%%%%%%%%%%%%%%%%

\section{\label{secthr}Threshold Behaviour}

The constraints on the threshold behaviour  of $\Pi(q^2)$ originate from 
our knowledge about the non-relativistic Greens function in the presence
of a Coulomb potential and its interplay with ``hard'' vertex
corrections. For a theory with non-vanishing $\beta$ function
(QED with light fermions or QCD) the proper definition of the
coupling constant and its running  must
be taken into account.

For completeness let us recall the ${\cal O}(\alpha_s)$ result.
To this order $R(s)$ is known analytically and reads 
(expressed in the variable $v=\sqrt{1-4m^2/s}$):
\begin{eqnarray}
R^{(0)} &=& 3 v \frac{3-v^2}{2},
\\
R^{(1)} &=& 3\frac{3-v^2}{2}
   \Bigg[
%%%\nonumber\\
%%%&&
    (1+v^2) \left(
            \mbox{Li}_2\left(\left(\frac{1-v}{1+v}\right)^2\right)
          +2\mbox{Li}_2\left(\frac{1-v}{1+v}\right)
          + \ln\frac{1+v}{1-v}\ln\frac{(1+v)^3}{8v^2}
                  \right)
\nonumber\\
&&
\hphantom{3\frac{3-v^2}{2}}
          + 3 v \ln\frac{1-v^2}{4v} - v\ln v
          + \frac{33+22v^2-7v^4}{8(3-v^2)}\ln\frac{1+v}{1-v}
          + \frac{15v-9v^3}{4(3-v^2)}
   \Bigg],
\label{r1exact}
\\
R(s) &=& R^{(0)} + C_F\frac{\alpha_s(s)}{\pi} R^{(1)} + {\cal O}(\alpha_s^2)
\label{r1ofsexact}
\\
     &=& 3 \frac{3}{2}v\left(1+\frac{\alpha_s}{\pi}C_F
                             \left(\frac{\pi^2(1+v^2)}{2v}-4\right)
                            +{\cal O}(v,\alpha_s^2)
                     \right).
\label{r1ofs}
\end{eqnarray}
In the last line an expansion for small velocities $v$ is performed.
The $\pi^2/2v$ term results from longitudinal gluon exchange,
the potential,
whereas the ``$-4$'' originates from a transversal, hard gluon.

For the present problem this has to be extended to ${\cal O}(\alpha_s^2)$.
Let us in a first step discuss the $C_F^2$ terms which are 
relevant for QCD and QED. 
For convenience everything will be formulated for QED with the
trivial translation $\alpha\to C_F\alpha_s$ implied for QCD.
The leading terms in an expansion of 
$x=\pi\alpha/v$ are given by
\begin{eqnarray}
R^v &=& \frac{3}{2}\frac{v x}{1-e^{-x}} 
     = \frac{3}{2}v\left(1+\frac{x}{2}
                          +\frac{B_1 x^2}{2!}
                          -\frac{B_2 x^4}{4!}
                          +\frac{B_3 x^6}{6!}
                          \pm\ldots
                          +(-1)^{(n+1)}\frac{B_n x^{2n}}{(2n)!}
                          \pm\ldots
                   \right)
\nonumber\\
    &=& \frac{3}{2}v\left(
       1 + \frac{\alpha}{\pi} \frac{\pi^2}{2v}
         + \left( \frac{\alpha}{\pi} \right)^2 \frac{\pi^4}{12 v^2}
         + \cdots
       \right)
\end{eqnarray}
where $B_n$ are the Bernoulli numbers: 
$B_1=1/6,\, B_2=1/30,\, B_3=1/42,\,\ldots$.
These terms are induced by 
rescattering through the Coulomb potential. Formally 
they are obtained from the imaginary part of the Greens function.
In addition there is a correction factor $(1-4\alpha/\pi)$ from 
transversal photon (gluon) exchange.
In QCD this term is quite familiar from the corrections to
quarkonium annihilation through a virtual photon
\cite{BarGatKoeKun75}.
These considerations
allow to reconstruct 
the $1/v$ and the constant term in order $\alpha^2$
\cite{BaiBro95,VolSmi94}
\begin{eqnarray}
R_A^{(2)} &=& 3\left(\frac{\pi^4}{8v} - 3\pi^2 + \ldots\right).
\label{rathr}
\end{eqnarray}
Now it is possible to define the leading term of $\Pi^{(2)}_{A}$ in such a
way that 
its imaginary part reproduces Eq.~(\ref{rathr}):
\begin{eqnarray}
\Pi^{(2)}_{A} &=& \frac{3}{16\pi^2}
            \left( \frac{\pi^5}{6\sqrt{1-z}} + 4\pi^2\ln(1-z)
                        +\ldots \right)
\end{eqnarray}
In Ref. \cite{BaiBro95} this information has been included by
considering the combination 
$\Pi_A^{(2)} + 4\Pi^{(1)}$
which is free from a logarithmic singularity at $q^2=4m^2$
and we shall adopt the same strategy.

In the discussion of the terms proportional to $C_F T n_l$
and $C_A C_F$ additional aspects must be taken into consideration
if one wants to predict the term of ${\cal O}(\alpha_s^2 v^0)$
multiplied by powers of logarithms.
This is related to the fact that for a prediction of 
${\cal O}(\alpha_s^2)$ the definition of the coupling constant, the 
renormalization scheme and the choice of scale affect the functional 
dependence of the result. The problem arises already in a
discussion of QED with additional $n_l$ light fermions.
Including purely photonic corrections up 
to order $\alpha$ and contributions with 
$n_l$ light fermion loops of order $\alpha^2$ one obtains in
QED close to threshold 
\cite{HoaKueTeu95}
\begin{eqnarray}
\!R_{QED}^{thr}
 \!\!&\!=\!&\!\! 
\frac{v(3-v^2)}{2} \left[1 
              + \frac{\bar{\alpha}(\mu^2)}{\pi}
                      \frac{\pi^2(1+v^2)}{2v}
              + n_l\left(\frac{\bar{\alpha}(\mu^2)}{\pi}\right)^2
                   \frac{\pi^2(1+v^2)}{6v}\left(
                                        \ln\frac{v^2 s}{\mu^2}
                                       -\frac{5}{3}
                                  \right)\right].
\label{cfnfthr}
\end{eqnarray}
where the $\overline{\mbox{MS}}$ definition of the coupling constant
has been adopted and the mass of the light fermions is set
to zero. This result is easily interpreted in the context of the QED
potential which can be written in the form
($\vec{q}$ denotes the relative momentum, with $\vec{q}=2\vec{p}$ in the
case at hand):
\begin{eqnarray}
V_{QED} & = & - 4\pi\frac{\alpha_V^{QED}(\vec{q}\,^2)}{\vec{q}\,^2},
\label{potentialqed}
\\
\alpha_V^{QED}(\vec{q}\,^2) &=& \bar{\alpha}(\mu^2) \left(
                      1 + n_l \frac{\bar{\alpha}(\mu^2)}{\pi}
                          \frac{1}{3}\left(
                               \ln\frac{\vec{q}\,^2}{\mu^2}-\frac{5}{3}
                                     \right)
                                                \right),
\label{alphaqed}
\end{eqnarray}
such that
\begin{equation}
R^{thr}_{QED}= \frac{v(3-v^2)}{2} \left[1
                 +\frac{\alpha_V(4\vec{p}\,^2)}{\pi}\frac{\pi^2(1+v^2)}{2v}
                                       \right].
\end{equation}

Returning to the $\overline{\mbox{MS}}$ convention and QCD
one now reconstructs the leading terms of the real part of 
$\Pi(q^2)$ close to threshold\footnote{
   The inclusion of the terms proportional to $v^2\ln v$ and 
   $v^4\ln v$ in addition to those proportional to $\ln v$
   in the ansatz will lead to a significant stabilization
   and improvement of our prediction.}
\begin{eqnarray}
\Pi_{l}^{(2),thr} &=& \frac{3}{16\pi^2}
\frac{\pi^2}{6}
                \left[1 + \frac{2}{3}\left(1-\frac{1}{z}\right)
                       - \frac{1}{3}\left(1-\frac{1}{z}\right)^2
                \right] 
\nonumber\\
&&
\hphantom{\frac{\pi^2}{6}}
      \left[-\ln^2\frac{1}{1-z}
            +2\ln\frac{1}{1-z}\left(\ln\frac{4m^2}{\mu^2}-\frac{5}{3}
                              \right)
      \right]
   + c_l 
\label{pilthr}
\end{eqnarray}
with an unknown constant term $c_l$.
This approach can immediately be transferred to QCD, replacing in 
Eq.~(\ref{potentialqed}) $\alpha_V^{QED}(\vec{q}\,^2)$ by 
$C_F\alpha_V^{QCD}(4\vec{p}\,^2)$ and using Eq.~(\ref{alphav}).
Thus for the $C_A C_F$ contribution we postulate the following 
threshold terms:
\begin{eqnarray}
R_{\it NA}^{(2),thr}&=&3\frac{\pi^2}{3}(3-v^2)(1+v^2)
                         \left(
                              -\frac{11}{16}\ln\frac{v^2 s}{\mu^2}
                              +\frac{31}{48}
                         \right),
\label{Rna}
\\
\Pi_{\it NA}^{(2),thr} &=&  \frac{3}{16\pi^2}
\frac{\pi^2}{6} 
                \left[1 + \frac{2}{3}\left(1-\frac{1}{z}\right)
                       - \frac{1}{3}\left(1-\frac{1}{z}\right)^2
                \right] 
\nonumber\\
&&
\hphantom{\frac{\pi^2}{6}}
      \left[\frac{11}{4}\ln^2\frac{1}{1-z}
            +8\ln\frac{1}{1-z}\left(
                                   -\frac{11}{16}\ln\frac{4m^2}{\mu^2}
                                   +\frac{31}{48}\right)
      \right]
    + c_{\it NA}. 
\label{pinathr}
\end{eqnarray}

Because $R_l^{(2)}$ and $R_{\it NA}^{(2)}$ develop only a logarithmic
singularity for $v\to 0$ it is not possible to determine the
leading real part completely, i.e. $c_l$ and $c_{\it NA}$ in
(\ref{pilthr}) and (\ref{pinathr}) remain unknown. 

%%%%%%%%%%%%%%%%%%%%%%%%%%%%%%%%%%%%%%%%%%%%%%%%%%%%%%%%%%%%
%%%%%%%%%%%%%%%%%%%%%%%%%%%%%%%%%%%%%%%%%%%%%%%%%%%%%%%%%%%%

\mathversion{bold}
\section {\label{secsma}Small $q^2$ Behaviour}
\mathversion{normal}

The first seven coefficients of the Taylor series around $q^2=0$
are calculated with the help of the program MATAD which has been used 
previously for the three-loop corrections to the $\rho$ parameter
\cite{CheKueSte95rho},
to $\Delta r$
\cite{CheKueSte95deltar}
and the top loop induced reaction $e^+e^- \to H l^+ l^-$
\cite{KniSte95}.
The package is written in FORM 
\cite{VerFORM}.
For a given diagram it performs the derivatives with respect to
the external momentum $q^2$ to the desired order, raising internal
propagators to powers up to about 20. It then performs the
traces and uses recurrence relations 
based on the integration by parts method
\cite{CheTka81,Bro92}
to reduce the resulting three-loop tadpole diagrams 
(in dimensional regularisation).
We find
\begin{eqnarray}
\Pi^{(0)} &=& \frac{3}{16\pi^2}\left(
      \frac{16}{15}z + \frac{16}{35}z^2 + \frac{256}{945}z^3 
    + \frac{128}{693}z^4 + \frac{2048}{15015}z^5 + \frac{2048}{19305}z^6 
    + \frac{65536}{765765}z^7
    + \ldots\right),
\nonumber\\ 
\Pi^{(1)} &=& \frac{3}{16\pi^2}\left(
      \frac{328}{81}z + \frac{1796}{675}z^2 + \frac{999664}{496125}z^3  
    + \frac{207944}{127575}z^4 + \frac{1729540864}{1260653625}z^5 
\right.
\nonumber\\
&+&
\left.
      \frac{21660988864}{18261468225}z^6
    + \frac{401009026048}{383490832725}z^7
    + \ldots \right),
\nonumber\\ 
\Pi^{(2)} &=& 
              \frac{3}{16\pi^2}
              \sum_{n>0} C_{n}^{(2)} z^n,
\\
C_{1}^{(2)} &=& C_F^2\left(
  -\frac{8687}{864}   + 4\zeta(2) - \frac{32}{5}\zeta(2)\ln2  
  + \frac{22781}{1728}\zeta(3)
               \right)
   +C_A C_F\left(
     \frac{127}{192} + \frac{902}{243}\lmm
\right.
\nonumber\\
&-& 
\left.
     \frac{16}{15}\zeta(2) 
   + \frac{16}{5}\zeta(2)\ln2
   + \frac{1451}{384}\zeta(3)
         \right)
   +C_F T n_l\left(
   - \frac{142}{243} - \frac{328}{243}\lmm - \frac{16}{15}\zeta(2)
             \right)
\nonumber\\
&+&
    C_F T \left(- \frac{11407}{2916} 
                 - \frac{328}{243}\lmm
                 + \frac{32}{15}\zeta(2) 
                 + \frac{203}{216}\zeta(3)
           \right),
\nonumber\\
C_{2}^{(2)} &=& C_F^2\left(
  - \frac{223404289}{1866240} 
  + \frac{24}{7}\zeta(2) - \frac{192}{35}\zeta(2)\ln2
  + \frac{4857587}{46080}\zeta(3) 
               \right)
\nonumber\\
&+& C_A C_F\left(
  - \frac{1030213543}{93312000} 
  + \frac{4939}{2025}\lmm
  - \frac{32}{35}\zeta(2) + \frac{96}{35}\zeta(2)\ln2
  + \frac{723515}{55296}\zeta(3) 
         \right)
\nonumber\\
&+& C_F T n_l\left(
  - \frac{40703}{60750} - \frac{1796}{2025}\lmm
  - \frac{32}{35}\zeta(2)
             \right)
\nonumber\\
&+& C_F T\left(-\frac{1520789}{414720}
             - \frac{1796}{2025}\lmm
             + \frac{64}{35}\zeta(2) 
             + \frac{14203}{18432}\zeta(3)
        \right),
\nonumber\\
C_{3}^{(2)} &=& C_F^2\left(
  - \frac{885937890461}{1161216000} 
  + \frac{64}{21}\zeta(2)
  - \frac{512}{105}\zeta(2)\ln2
  + \frac{33067024499}{51609600}\zeta(3) 
               \right)
\nonumber\\
&+& C_A C_F\left( 
  - \frac{95905830011197}{1706987520000} 
  + \frac{2749076}{1488375}\lmm 
  - \frac{256}{315}\zeta(2)
  + \frac{256}{105}\zeta(2)\ln2
         \right.
\nonumber\\
  &+& \left. \frac{5164056461}{103219200}\zeta(3) 
         \right)
+ C_F T n_l\left(
  - \frac{9703588}{17364375} - \frac{999664}{1488375}\lmm 
  - \frac{256}{315}\zeta(2)
             \right)
\nonumber\\
&+& C_F T \left(
       -\frac{83936527}{23328000}
       -\frac{999664}{1488375}\lmm 
       +\frac{512}{315}\zeta(2) 
       +\frac{12355}{13824}\zeta(3)
             \right),
\nonumber\\
C_{4}^{(2)} &=& C_F^2 \left(
  - \frac{269240669884818833}{61451550720000}
  + \frac{640}{231}\zeta(2) - \frac{1024}{231}\zeta(2)\ln2 
  + \frac{1507351507033}{412876800}\zeta(3)
                \right)
\nonumber\\
&+& C_A C_F\left(
   - \frac{36675392331131681}{158018273280000}
   + \frac{571846}{382725}\lmm 
   - \frac{512}{693}\zeta(2) + \frac{512}{231}\zeta(2)\ln2
         \right.
\nonumber\\
   &+& \left.
       \frac{1455887207647}{7431782400}\zeta(3)
         \right)
+ C_F T n_l\left(
  - \frac{54924808}{120558375} - \frac{207944}{382725}\lmm 
  - \frac{512}{693}\zeta(2)
             \right)
\nonumber\\
&+& C_F T \left(
          -\frac{129586264289}{35831808000}
          -\frac{207944}{382725}\lmm 
          +\frac{1024}{693}\zeta(2) 
          +\frac{2522821}{2359296}\zeta(3)
          \right),
\nonumber\\
C_{5}^{(2)} &=& C_F^2 \left(
         -\frac{360248170450504167133}{15209258803200000} 
         +\frac{2560}{1001}\zeta(2)
         -\frac{4096}{1001}\zeta(2)\ln2 
\right.
\nonumber\\
&+&
\left.
         \frac{939939943788973}{47687270400}\zeta(3)
                \right)
%%%\nonumber\\
%%%&+&  
+C_A C_F\left(
       -\frac{21883348499544169357}{23658847027200000} 
\right.
\nonumber\\
&+&
\left.
        \frac{432385216}{343814625}\lmm 
       -\frac{2048}{3003}\zeta(2) 
       +\frac{2048}{1001}\zeta(2)\ln2
       +\frac{14724562345079}{19074908160}\zeta(3)
            \right) 
\nonumber\\
&+& 
 C_F T n_l\left(
        -\frac{4881989801536}{13104494431875} 
        -\frac{1729540864}{3781960875}\lmm 
        -\frac{2048}{3003}\zeta(2)
             \right)
\nonumber\\
&+& 
 C_F T  \left( 
          -\frac{512847330943}{139087872000}
          -\frac{1729540864}{3781960875}\lmm 
          +\frac{4096}{3003}\zeta(2)
          +\frac{1239683}{983040}\zeta(3)
            \right),
\nonumber\\
C_{6}^{(2)} &=& C_F^2 \left(
        -\frac{64959156551995419148501103}{529285210649395200000} 
        +\frac{1024}{429}\zeta(2) 
        -\frac{8192}{2145}\zeta(2)\ln2
\right.
\nonumber\\
&+&
\left.
         \frac{330704075360938001}{3238841548800}\zeta(3)
                 \right) 
%%%\nonumber\\
%%%&+& 
+ C_A C_F\left(
       -\frac{4826864658245605658856745531}{1317342772469012889600000} 
\right.
\nonumber\\
&+&
\left.
       +\frac{5415247216}{4980400425}\lmm 
       -\frac{4096}{6435}\zeta(2) 
       +\frac{4096}{2145}\zeta(2)\ln2
       +\frac{580922571682067161}{190443883069440}\zeta(3)
            \right) 
\nonumber\\
&+& 
 C_F T n_l\left(
        -\frac{151249070952032}{493552701717075} 
        -\frac{21660988864}{54784404675}\lmm 
        -\frac{4096}{6435}\zeta(2)
          \right)
\nonumber\\
&+& 
 C_F T \left(
        -\frac{3411069430668887863}{899847347503104000}
        -\frac{21660988864}{54784404675}\lmm 
        +\frac{8192}{6435}\zeta(2)
\right.
\nonumber\\
&+& 
\left.
         \frac{1760922667}{1207959552}\zeta(3)
          \right),
\nonumber\\
C_{7}^{(2)} &=& C_F^2 \left(
        -\frac{571365897351090627148045413923471}
              {927409311818185074278400000} 
        +\frac{16384}{7293}\zeta(2) 
        -\frac{131072}{36465}\zeta(2)\ln2  
\right.
\nonumber\\
&+&
\left.
         \frac{13386367971827490465799}{26118018249523200}\zeta(3)
                 \right) 
\nonumber\\
&+& 
    C_A C_F\left(
       -\frac{7342721436809271685822267340249}{505859624628100949606400000} 
       +\frac{100252256512}{104588408925}\lmm 
\right.
\nonumber\\
&-&
\left.
        \frac{65536}{109395}\zeta(2) 
       +\frac{65536}{36465}\zeta(2)\ln2
       +\frac{14019414333929589373}{1160800811089920}\zeta(3) 
            \right) 
\nonumber\\
&+& 
 C_F T n_l\left(
        -\frac{13125091764358528}{51823033680292875} 
        -\frac{401009026048}{1150472498175}\lmm 
        -\frac{65536}{109395}\zeta(2) 
          \right)
\nonumber\\
&+& 
 C_F T \left(
        -\frac{7927736038867601807}{2024656531881984000}
        -\frac{401009026048}{1150472498175}\lmm 
        +\frac{131072}{109395}\zeta(2) 
\right.
\nonumber\\
&+& 
\left.
         \frac{4497899939}{2717908992}\zeta(3)
          \right).
\nonumber
\end{eqnarray}
The coefficients $C_1^{(2)} - C_4^{(2)}$ were already presented in
\cite{CheKueSte95Pade}, $C_5^{(2)}, C_6^{(2)}$ and $C_7^{(2)}$ are new.
The $C_F^2$ term of $C_1^{(2)} - C_3^{(2)}$ and $C_5^{(2)}$ are 
in agreement with 
\cite{BaiBro95} and \cite{Bai96}, respectively.

The $\mu$ dependence of the terms proportional to $C_A C_F$ and $C_F T n_l$
is again due to the running of $\alpha_s(\mu^2)$. If the 
$\overline{\mbox{MS}}$ definition of the mass is used, the Abelian part
also develops a $\mu$ dependence --- a consequence of the anomalous mass
dimension --- and in general $\ln^2\mu^2$ terms appear.
The corresponding expressions can easily be derived using 
Eq.~(\ref{msmass2osmass}) and will not be listed.
Instead we want to apply the BLM procedure
\cite{BroLepMac83}
to our results. 
\begin{table}[hb]
\renewcommand{\arraystretch}{1.3}
\begin{center}
\begin{tabular}{|l||r|r||r|r|r||r|r|r||r|r|}
%%%\hline
\hline
n & $C_n^{(0)}$ & $C_n^{(1)}$ 
  & $C_{A,n}^{(2)}$ & $C_{{\it NA},n}^{(2)}$ & $C_{l,n}^{(2)}$ 
  & $C_n^{(2)}$ 
  & $C_n^{BLM}$ & $\mu_{BLM}/m$ 
  & $\bar{C}_n^{(1)}$ & $\bar{C}_n^{(2)}$ \\
\hline
\hline
1 & 1.07 & 4.05 & 5.08 & 7.10 & -2.34 & 30.1 & 13.7 & 0.420 & 1.92 & 3.82 \\
2 & 0.46 & 2.66 & 6.39 & 6.31 & -2.17 & 29.5 & 14.3 & 0.294 & 0.83 & 3.69 \\
3 & 0.27 & 2.01 & 6.69 & 5.40 & -1.90 & 27.3 & 14.0 & 0.244 & 0.39 & 2.50 \\
4 & 0.18 & 1.63 & 6.68 & 4.70 & -1.67 & 25.2 & 13.5 & 0.215 & 0.15 & 1.65 \\
5 & 0.14 & 1.37 & 6.57 & 4.16 & -1.49 & 23.4 & 13.0 & 0.195 & 0.01 & 1.12 \\
6 & 0.11 & 1.19 & 6.43 & 3.75 & -1.35 & 21.9 & 12.5 & 0.181 &-0.09 & 0.85 \\
7 & 0.09 & 1.05 & 6.27 & 3.41 & -1.24 & 20.7 & 12.0 & 0.169 &-0.15 & 0.76 \\
\hline
%%%\hline
\end{tabular}
\end{center}
\caption{\label{tabblm} The BLM scale $\mu_{BLM}$ is expressed in terms of the 
                        original scale $m$. For the numerical values of 
                        $C_n^{(2)}$ and $\bar{C}_n^{(2)}$ 
                        $n_l=5$ and $\mu^2=m^2$, respectively, 
                        $\mu^2=\bar{m}^2$  
                        is used. The double bubble
                        diagrams with two massive fermion loops are also
                        included. $C_n^{BLM}$ is the coefficient remaining
                        after the $\beta_0$ term is absorbed.}
\end{table}
It suggests to choose the scale of the
${\cal O}(\alpha_s)$ term in such a way that the contribution of the 
${\cal O}(\alpha_s^2)$ part proportional to $\beta_0$ is absorbed.
This prescription is based on the observation that the remaining coefficients
of the $\alpha_s^2$ terms are often relatively small.
It is possible to treat each term of ${\cal O}(z^n)$
separately. In Table \ref{tabblm} we list the BLM scale
$\mu_{BLM}$ for each Taylor coefficient together with the
numerical value of the original ($C_n^{(2)}$) and the
${\cal O}(\alpha_s^2)$ term which remains after the BLM scale 
is adjusted ($C_n^{BLM}$).
It is interesting to note that $\mu_{BLM}$
is decreasing with increasing $n$. 
This is plausible because the higher Taylor coefficients
are increasingly dominated by
the threshold region and the characteristic scale at
threshold is the relative three-momentum of the quarks.
Note that $C_n^{BLM}$ remains nearly constant whereas $C_n^{(2)}$ 
decreases for increasing $n$ (but remember the rapidly decreasing 
coefficients $C_n^{(0)}$). 

In comparison to $C_n^{(2)}$ in Table \ref{tabblm} also the
corresponding one- and two-loop coefficients ($C_n^{(0)}, C_n^{(1)}$)
and the two- and three-loop coefficients ($\bar{C}_n^{(1)}, \bar{C}_n^{(2)}$)
in the $\overline{\mbox{MS}}$ scheme are listed. Whereas
$C_n^{(1)}$ is roughly of ${\cal O}(1)$ 
$\bar{C}_n^{(1)}$ varies from approximately $1.9$ down to $0.01$.
Also the sign changes. Therefore the BLM procedure is rather unstable and 
not applicable. On the other hand, the 
$\overline{\mbox{MS}}$ coefficients $\bar{C}_n^{(2)}$ are 
already reasonably small, so there is no urgent need for an optimization
procedure. 

Using Eq.~(\ref{alphav}) it is possible to transform
$\alpha_s(\mu^2)$ 
into $\alpha_V(\mu^2)$, which is based on a physical definition.
We observed that the quantities $C_n^V$ are already smaller than
the corresponding on-shell ones 
(but still much larger than in the $\overline{\mbox{MS}}$ scheme).
After application of the BLM prescription, however, the
coefficients are larger than the corresponding $C_n^{BLM}$.
No significant reduction of the coefficients is achieved.

%%%%%%%%%%%%%%%%%%%%%%%%%%%%%%%%%%%%%%%%%%%%%%%%%%%%%%%%%%%%
%%%%%%%%%%%%%%%%%%%%%%%%%%%%%%%%%%%%%%%%%%%%%%%%%%%%%%%%%%%%

\mathversion{bold}
\section{The QED Coupling Constant in the $\overline{\mbox{MS}}$ Scheme}
\label{secQED}
\mathversion{normal}

The vacuum polarization is connected with the physical cross section
through the dispersion relation:
\begin{eqnarray}
\Pi(q^2) &=& \frac{q^2}{12\pi^2} \int_{4m^2}^\infty \,ds\,\, 
                \frac{R(s)}{s(s-q^2)} + \frac{3}{16\pi^2}\,C_0.
\label{eqdisprel}
\end{eqnarray}
As is well known $\Pi(q^2)$
by itself is not a physical quantity. This is reflected  by the presence
of an arbitrary subtraction constant $C_0$ in  Eq.~(\ref{eqdisprel}).
In QED usually the condition
\begin{eqnarray}
\Pi(0) &=& 0 
\end{eqnarray}
is adopted
which  obviously leads to $C_0=0$. Then the  photon propagator 
\begin{eqnarray}
D(q) &=& \frac{1}{q^2}\frac{1}{1 + e^2\Pi(q^2)}
\end{eqnarray}
gets canonically normalized in the vicinity of the photon pole
$q^2  \approx 0$ and the electromagnetic charge $e$ 
coincides with the one defined in the Thompson limit. 
On the other hand, using the $\overline{\mbox{MS}}$ scheme
to renormalize the vacuum polarization function, 
the photon propagator looks like
\begin{eqnarray}
\bar{D}(q) &=& \frac{1}{q^2}
               \frac{1}{1 + \bar{e}^2 \bar{\Pi}(q^2)},
\end{eqnarray}
which leads to
\begin{eqnarray}
\alpha  &=&   \frac{\bar{\alpha}}
                        {1 + \bar{e}^2 \bar{\Pi}(q^2=0)} 
        \,=\,  \frac{\bar{\alpha}}{1 
       + \frac{3\bar{\alpha}}{4\pi}\bar{C}_0}.
\end{eqnarray}
This relation is not well defined if one takes into account massless 
quark contributions to $\Pi$ (even within perturbation theory),
but it correctly describes the part of the relation between on-shell and
$\overline{\mbox{MS}}$ electromagnetic coupling constant
coming from a massive quark coupled to the electromagnetic 
current. For a sufficiently heavy quark this relation could even be 
used for a reliable theoretical estimation of the quark contribution
to the charge renormalization.
The Taylor series for the vacuum polarization discussed in the
previous section was evaluated in a first step in the 
$\overline{\mbox{MS}}$ scheme, with a constant term given by
\begin{eqnarray}
\lefteqn{\frac{16\pi^2}{3} \bar{\Pi}(0) \,\,=\,\, \bar{C}_0 \,\,=\,\,} 
\nonumber\\
&& 
       \frac{4}{3}\lmm 
     + \frac{\alpha_s(\mu^2)}{\pi} C_F \left(\frac{15}{4} + \lmm\right) 
     + \left(\frac{\alpha_s(\mu^2)}{\pi}\right)^2\Bigg[
\nonumber\\
&&
       C_F^2 \left( \frac{77}{144} 
                   - \frac{1}{8}\lmm 
                   + 5\zeta(2) 
                   - 8\zeta(2)\ln2 
                   + \frac{1}{48}\zeta(3) 
              \right) 
\nonumber\\
&&
      + C_AC_F\left( \frac{14977}{2592} 
                   + \frac{157}{36}\lmm 
                   + \frac{11}{24}\lmmz 
                   - \frac{4}{3}\zeta(2)
                   + 4\zeta(2)\ln2 
                   + \frac{127}{96}\zeta(3) 
              \right)
\nonumber\\
&&
      + C_FTN_f\left(-\frac{917}{648} 
                    - \frac{14}{9}\lmm 
                    - \frac{1}{6}\lmmz 
                    - \frac{4}{3}\zeta(2)
               \right)
\nonumber\\
&&
      + C_FT \left(-\frac{23}{8} 
                  + 4\zeta(2) 
                  + \frac{7}{16}\zeta(3) 
             \right)
                                                 \Bigg].
\end{eqnarray}
This leads to the following relation between $\alpha$
and $\bar{\alpha}(\mu^2)$:
\begin{eqnarray}
\alpha &=& \bar{\alpha}(\mu^2)\left[1 
                     - \frac{3\bar{\alpha}(\mu^2)}{4\pi}\bar{C}_0 \right],
\\
\bar{\alpha}(\mu^2) &=& \alpha\left[1 + \frac{3\alpha}{4\pi}\bar{C}_0 \right].
\end{eqnarray}
For the QED case (i.e. $C_F=1, C_A=0, T=1/2$ and $N_f=1$)
the result of \cite{Bro92} is reproduced.

%%%%%%%%%%%%%%%%%%%%%%%%%%%%%%%%%%%%%%%%%%%%%%%%%%%%%%%%%%%%
%%%%%%%%%%%%%%%%%%%%%%%%%%%%%%%%%%%%%%%%%%%%%%%%%%%%%%%%%%%%

\section{\label{secoptappr}Optimized Approximation}

It is evidently a promising strategy to construct, in a first step, a
function $\tilde{\Pi}(q^2)$ 
whose discontinuity across the cut 
approximates the one of the full vacuum polarization up to subleading 
terms --- in other words, which exhibits a similar spectral function. 
The approximation will have different structure for 
$\Pi_A$ on one side and 
$\Pi_{\it NA}$ and
$\Pi_l$ 
on the other side as a consequence of their different threshold 
singularities. 

For the following discussion the on-shell scheme for the mass $m$
is used and the 't~Hooft mass $\mu^2$ is set to $m^2$.
For the $C_F^2$ term we proceed as in 
\cite{BaiBro95}
and define $\tilde{\Pi}_A$ as follows:
\begin{eqnarray}
\tilde{\Pi}_A^{(2)}(q^2) &=&
   \Pi_A^{(2)}(q^2) + 4\Pi^{(1)}(q^2)
\nonumber\\
&& +\frac{3}{16\pi^2}\Bigg[
    (1-z) G(z)\left(9 G(z) + \frac{229+62z}{8z} \right)
    -\frac{229}{8z}
    -\frac{173}{24}
                    \Bigg].
\label{pofwa}
\end{eqnarray}
As already mentioned above the combination 
$\Pi_A^{(2)}(q^2) + 4\Pi^{(1)}(q^2)$ is free of logarithmic
singularities at threshold whereas the $1/v$ term remains and is
treated below.
The high energy logarithms have to be removed in such a way
that the threshold behaviour and the polynomial structure
for $q^2\to0$ is not destroyed. Therefore we use the function
$G(z)$ which is essentially the result in the one-loop
case. The terms quadratic in $G(z)$ remove the $1/z\ln^2(-z)$ terms
whereas the linear ones develop only single logarithms for 
$q^2\to -\infty$. 
The factor $(1-z)$ ensures that the threshold is not touched.
The last two terms reestablish the condition that the new 
polarization function also vanishes for $q^2=0$.

\begin{figure}[ht]
 \begin{center}
 \begin{tabular}{c}
   \epsfxsize=11.5cm
   \leavevmode
   \epsffile[80 280 540 520]{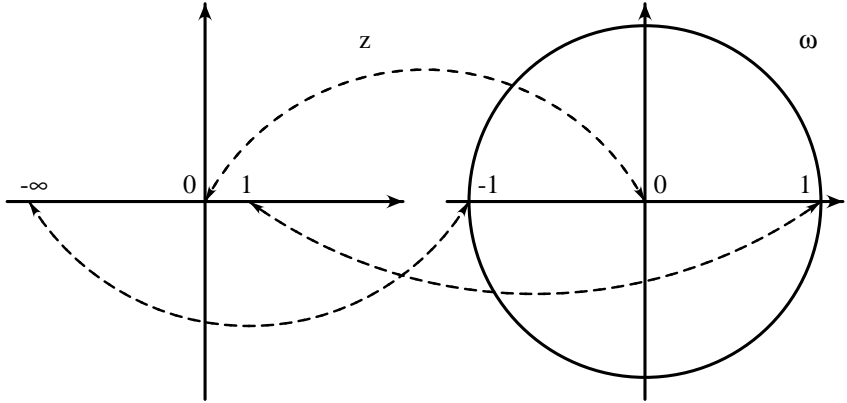}
 \end{tabular}
 \caption{\label{complexplane} Transformation between the 
                              $z$ and $\omega$ plane }
 \end{center}
\end{figure}

In a next step we map the complex $q^2$-plane into the interior
of the unit circle with the help of the variable transformation
\cite{FleCM}
\begin{eqnarray}
\omega = \frac{1-\sqrt{1-q^2/4m^2}}{1+\sqrt{1-q^2/4m^2}},\,\,\,\,
&&
\frac{q^2}{4m^2} = \frac{4\omega}{(1+\omega)^2}.
\label{omega}
\end{eqnarray}
The special points
$q^2=0,4m^2,-\infty$ correspond to $\omega=0,1,-1$, respectively.
In Figure~\ref{complexplane} this conformal mapping is
visualized. The upper (lower) part of the cut form $z=1$ to $\infty$
is mapped onto the upper (lower) perimeter of the circle. The
Taylor series which in the variable $z$ only converges
for $|q^2|<4m^2$ is now convergent in the interior of the
unit circle in the complex $\omega$ plane.

To improve the approximation 
in a way which allows to approach the boundary of the domain
of analycity
it has been suggested to use
Pad\'e approximations. Following
\cite{BaiBro95}
we define
\begin{eqnarray}
P_A(\omega) &=& \frac{1-\omega}{(1+\omega)^2}
\left[\tilde{\Pi}_A^{(2)}(q^2) - \tilde{\Pi}_A^{(2)}(-\infty)\right].
\end{eqnarray}
The knowledge of $\tilde{\Pi}_A^{(2)}(q^2)$ in the three different kinematical
regions translates into information for $P(\omega)$ for 
$\omega=-1,0$ and $1$.
Due to the fact that $P^\prime(-1)=0$ (this is because of the definition 
of $\omega$) the constant term in the high energy limit, 
$\tilde{\Pi}_A^{(2)}(-\infty)$, is subtracted.  
The factor $1/(1+\omega)^2$ then guarantees that in the limit $\omega\to -1$
the power suppressed term of $\tilde{\Pi}_A^{(2)}(q^2)$
is projected out whereas the factor $(1-\omega)$ together with the 
$1/v$-singularity of $\tilde{\Pi}(q^2)$ results in a finite value for
$P(1)$.
For the construction of the Pad\'e approximants the following
information about $P_A(\omega)$ is available:
$\{P_A(-1),P_A(0),P_A^{\prime}(0),\ldots,P_A^{(7)}(0),P_A(1)\}$. These
ten data points are enough to construct Pad\'e approximants
like $[4/5]$ or $[5/4]$.

The above procedure has to be modified for
$\Pi_{\it NA}^{(2)}$ and $\Pi_l^{(2)}$ because of the different
threshold behaviour. Here the logarithmic singularities 
of $\Pi_{{\it NA}/l}$ are
subtracted before the Pad\'e approximation is performed.
Therefore $\tilde{\Pi}_{\it NA}^{(2)}$ and $\tilde{\Pi}_l^{(2)}$
looks like
\begin{eqnarray}
\tilde{\Pi}_{\it NA}^{(2)}(q^2) &=& \Pi_{\it NA}^{(2)}(q^2) 
\nonumber\\
&&
  -\left[\Pi_{\it NA}^{(2),thr}(q^2) - c_{\it NA}
        +\frac{3}{16\pi^2}\frac{\pi^2}{6}
          \left(                 \frac{16}{9}  - \frac{11}{12}\ln4
               +\frac{1}{z}\left(\frac{31}{18} - \frac{11}{6}\ln4\right)
          \right)    
   \right]
%%%\right.
\nonumber\\
&&
%%%\left.
  +\frac{3}{16\pi^2}\Bigg[
    z(1-z) G(z) \left(
        \left( \frac{11}{2z} + \frac{11}{6} - \frac{44}{3}\zeta(2) \right)G(z)
\right.
\nonumber\\
&&
\left.
  +\frac{1413 + 738 z - \zeta(2)(1024 + 992 z) - \zeta(3)(264 +528 z)}{72z^2}
                \right)
\nonumber\\
&&      
      -\frac{157}{8z} -\frac{221}{24}
      +\zeta(2)\left(\frac{128}{9z}+\frac{244}{27}\right)
      +\zeta(3)\left(\frac{11}{3z}+\frac{55}{9}\right)
                    \Bigg],
\label{pismartna}
\\
\tilde{\Pi}_l^{(2)}(q^2) &=& \Pi_l^{(2)}(q^2) 
\nonumber\\
&&
- \left[\Pi_l^{(2),thr}(q^2) - c_{\it l}
        +\frac{3}{16\pi^2}\frac{\pi^2}{6}
          \left(                 -\frac{8}{9}  + \frac{1}{3}\ln4
               +\frac{1}{z}\left(-\frac{10}{9} + \frac{2}{3}\ln4\right)
          \right)
  \right]
\nonumber\\
&&
  +\frac{3}{16\pi^2}\Bigg[
    z(1-z) G(z) \left( 
        \left(-\frac{2}{z}-\frac{2}{3}+\frac{16}{3}\zeta(2)\right) G(z)
\right.
\nonumber\\
&&
\left.
       +\frac{-105-66z+\zeta(2)(128+160z)+\zeta(3)(24+48z)}{18z^2}
                \right) 
\nonumber\\
&&
      +\frac{35}{6z}+\frac{67}{18}
       +\zeta(2)\left(-\frac{64}{9z}-\frac{176}{27}\right)
       +\zeta(3)\left(-\frac{4}{3z} -\frac{20}{9}\right)
                   \Bigg].
\end{eqnarray}
Again the function $G(z)$ is used to get rid of the high energy
logarithms. The additional factor $z$ in front of the
$(1-z)(G(z))^2$ part has to be introduced because of the 
$\ln^2 (-z)$ behaviour of
$\Pi_{\it NA}^{(2)}$ and $\Pi_l^{(2)}$ for $z\to -\infty$.
The terms next to $\Pi_{{\it NA}/l}^{(2),thr}$ are needed in order to ensure
that $\tilde{\Pi}_{{\it NA}/l}^{(2)}(0)=0$.
The mapping function is defined as 
\begin{eqnarray}
P_{{\it NA}/l}(\omega) &=& \frac{1}{(1+\omega)^2}
\left[\tilde{\Pi}_{{\it NA}/l}^{(2)}(q^2) 
    - \tilde{\Pi}_{{\it NA}/l}^{(2)}(-\infty)\right]
\label{pofwnal}
\end{eqnarray}
and besides $P_{{\it NA}/l}(-1)$ and $P_{{\it NA}/l}(0)$ the first seven
derivatives for $\omega=0$ are available. 
Because of the unknown constants $c_l$ and $c_{\it NA}$ defined in
(\ref{pilthr},\ref{pinathr}) $P(1)$ is not available.
This allows the construction
of Pad\'e approximants like $[4/4]$ or $[5/3]$.

For the non-Abelian part, $R_{\it NA}^{(2)}$, there is an other
possibility to construct an approximation. This second method
is based on the observation 
that for a special choice of the gauge parameter $\xi$
the complete threshold behaviour and the logarithmic divergency
for $s\to\infty$ is covered by the gluonic double bubble
diagrams given by the last two diagrams of Fig.~\ref{fig3loop}
(together with the corresponding ghost graphs).
Recently, exact results for these contributions with
arbitrary $\xi$ were presented
\cite{CheHoaKueSteTeu96} which are repeated in Appendix
\ref{appdouble}.
It is therefore a strong temptation to consider 
$\Pi_{\it NA}^{(2)} - \Pi_{g}^{(2)}|_{\xi=4}$ from the very
beginning. The imaginary part of this difference is 
already numerically small,
vanishes at threshold and has no logarithmic singularity
for $s\to\infty$.
In analogy to Eq.~(\ref{pismartna}) and (\ref{pofwnal}) we define
\begin{eqnarray}
\tilde{\Pi}_{\it NA}^{g,(2)}(q^2) &=& 
           \Pi_{\it NA}^{(2)}(q^2) - \Pi_{g}^{(2)}(q^2)\Big|_{\xi=4}
  +\frac{3}{16\pi^2}\Bigg[
                          (1-z) G(z) \frac{33+6z}{4z} 
                        - \frac{33}{4z}
                        + \frac{5}{4}
                    \Bigg],
\label{pismartnagldb}\\
P_{\it NA}^g(\omega) &=& \frac{1}{(1+\omega)^2}
\left[\tilde{\Pi}_{\it NA}^{g,(2)}(q^2) 
    - \tilde{\Pi}_{\it NA}^{g,(2)}(-\infty)\right]
\label{pofwnagldb}
\end{eqnarray}
and proceed in the same way as above.
The coefficients of $\Pi_{g}^{(2)}(q^2)|_{\xi=4}$ for $q^2\to0$ and the high
energy expansion can be calculated from $\Pi_l^{(2)}(q^2)$ using 
the formula
\cite{CheHoaKueSteTeu96}
(see also Appendix \ref{appdouble})
\begin{eqnarray}
\Pi_{g}^{(2)}(q^2)\Big|_{\xi=4} &=&- \frac{11}{4}\Pi_l^{(2)}(q^2) 
                         - \frac{2}{3}\Pi^{(1)}(q^2)
\end{eqnarray}
in the different kinematical regions.

We have checked that both methods lead to equivalent results. The
second method is, however, more appealing and the results presented below
are based on the second choice.

%%%%%%%%%%%%%%%%%%%%%%%%%%%%%%%%%%%%%%%%%%%%%%%%%%%%%%%%%%%%
%%%%%%%%%%%%%%%%%%%%%%%%%%%%%%%%%%%%%%%%%%%%%%%%%%%%%%%%%%%%

\section{\label{secres}Results}

After the Pad\'e approximation is performed the  
Eqs.~(\ref{pofwa},\ref{pofwnal},\ref{pofwnagldb})
are inverted and the threshold and high energy terms
are reintroduced. Thus real and imaginary parts of
$\Pi^{(2)}$ are at hand.
Let us subsequently discuss the absorptive part $R^{(2)}$.
We choose $\mu^2=m^2$ throughout the whole discussion.

Figure~\ref{figfull} shows the complete result for $R_A$,
$R_{\it NA}$ and $R_l$ plotted against the velocity $v$ (solid line).
This presentation expands the threshold region.
The dashed lines represent the high energy and threshold
approximations. The curves are dominated by the 
threshold singularities. In Figure~\ref{fighigh} the same 
results are shown with abscissa $x=2m/\sqrt{s}$. 
This presentation expands the high energy region.
In addition to
the quadratic approximations which are incorporated into 
$R^{(2)}$ also the quartic approximations 
(Eqs.~(\ref{rm4a}-\ref{rm4l}))
are plotted. They
are evidently very well reproduced by our method.
The exact result for $R_l$ from
\cite{HoaKueTeu95}
is indistinguishable from the 
curves in Figure~\ref{figfull} and \ref{fighigh}.
Also different Pad\'e approximants $[5/4], [4/5], [4/4], \ldots$
deliver practically identical results 
with a variation comparable to the thickness of the lines. 
Small variations are visible
close to threshold {\it after} subtraction of the singular and constant 
term. Figure~\ref{figthr} shows the results of up to ten different Pad\'e
approximants. The spread indicates the error of the method.
The agreement in the high energy region is even better.
In this Figure only those curves are plotted which have no poles
inside the unit circle in the $\omega$ plane.
Perfect agreement for $R_l^{(2)}$ is observed which is even slightly
increased if higher order Pad\'e approximants (solid curves)
are taken into account.  It is hard to detect the exact result which 
is represented by the dotted line.
$R_{\it NA}^{(2)}$ seems to converge to the solid line 
($[4/4], [5/3]$ and $[3/5]$)
when more moments from small $q^2$ are included. The dashed lines are from
the $[3/3], [4/2], [2/4]$ and $[3/4]$, the dotted
ones from Pad\'e approximants including only 
$C_{{\it NA},3}^{(2)}$ and $C_{{\it NA},4}^{(2)}$ 
($[3/1], [2/2]$ and $[3/2]$).
The dash-dotted curve is the [4/3] Pad\'e approximant and has a pole very 
close to $\omega=1 (1.07\ldots)$.
For the Abelian part a classification of the different results
can be seen: the dashed lines are $[4/2]$ and $[2/4]$,
the solid ones $[3/2], [2/3], [5/3], [3/5]$ and $[5/4]$ 
Pad\'e approximants.

In the following handy approximation formulae for
$R_A^{(2)}(s)$ and $R_{\it NA}^{(2)}(s)$
are presented. Details can be found in Appendix \ref{appapp}.
The results read:
\begin{eqnarray}
R_A^{(2)} &=& \frac{(1-v^2)^4}{v}\frac{3\pi^4}{8}
              - 4 R^{(1)}
                +v\frac{2619}{64}-v^3\frac{2061}{64}
                +\frac{81}{8}\left(1-v^2\right)\ln\frac{1-v}{1+v}
\nonumber\\
&& 
-198\left(\frac{m^2}{s}\right)^{3/2} \left(v^4-2v^2\right)^6 
\nonumber\\
&&
+100 p^{3/2} (1-p) \left[
2.21 P_0(p)
-1.57 P_1(p)
+0.27 P_2(p)
\right],
\label{appfora}
\\
R_{\it NA}^{(2)} &=& R_{g}^{(2)}\Big|_{\xi=4}
               + v\frac{351}{32} - v^3\frac{297}{32}
\nonumber\\
&& 
-18\left(\frac{m^2}{s}\right)^{3/2}  \left(v^4-2v^2\right)^4
\nonumber\\
&& 
+50 p^{3/2} (1-p) \left[
1.73 P_0(p)
-1.24 P_1(p)
+0.64 P_2(p)
\right],
\label{appfornaxi}
\end{eqnarray}
with $p=(1-v)/(1+v)$.
The first line of Eq.~(\ref{appfora}) and (\ref{appfornaxi})
consists of the exactly known high energy and
threshold contribution,
whereas the second and third
line represents the numerically small remainder
which is plotted in Figures~\ref{figrema} and \ref{figremnaxi},
respectively, 
together with the result from the Pad\'e approximation.

%%%%%%%%%%%%%%%%%%%%%%%%%%%%%%%%%%%%%%%%%%%%%%%%%%%%%%%%%%%%
%%%%%%%%%%%%%%%%%%%%%%%%%%%%%%%%%%%%%%%%%%%%%%%%%%%%%%%%%%%%

\section{\label{seccon}Conclusions and Summary}

Real and imaginary parts of the three-loop polarization of
heavy quarks have been calculated. These results can be considered
as natural extension of results by K\"all\'en and Sabry obtained
more than four decades ago. At an intermediate step
predictions for the moments (up to $n=7$) are obtained which
are basic ingredients for our numerical method, and may in addition
also be of interest for the evaluation of QCD sum rules
involving heavy quarks. The results for the imaginary part
allow to predict the cross section in order $\alpha_s^2$. The
Pad\'e approximation leads to fairly stable predictions
with a relative uncertainty for the $\alpha_s^2$ coefficients
of at most 5\%.
The prediction in its present form can be considered as reliable
down to fairly small values of $v$, say to values of $x=4/3\pi\alpha_s/v$
around $2$. For even smaller values of $v$ the leading terms 
of the expansion in $x$ have to be resummed. This aspect
will be treated in a future publication.

\vspace{5ex}                  
\noindent
{\bf Acknowledgments}

\noindent
We would like to thank A.H. Hoang and T. Teubner for 
many interesting discussions. Their programs with the
analytical results for 
$R_{\mbox{\scriptsize\it l}}^{(2)}$ were crucial for our tests
of the approximation methods. 
We are grateful to D. Broadhurst for helpful discussions.
Comments by S. Brodsky and M. Voloshin on the threshold behaviour
are gratefully acknowledged.

\vspace{5ex}                  
\noindent
{\Large \bf Appendix}

%%%%%%%%%%%%%%%%%%%%%%%%%%%%%%%%%%%%%%%%%%%%%%%%%%%%%%%%%%%%
%%%%%%%%%%%%%%%%%%%%%%%%%%%%%%%%%%%%%%%%%%%%%%%%%%%%%%%%%%%%

\renewcommand {\theequation}{\Alph{section}.\arabic{equation}}
\begin{appendix}

\setcounter{equation}{0}
\section{\label{appapp}Approximation Formulae} 

In this appendix handy approximations for
$R_A^{(2)}(s)$ and $R_{\it NA}^{(2)}(s)$
are derived. 
We write $R^{(2)}(s)$
for the different colour factors in the form:
\begin{eqnarray}
R_A^{(2)}(s) &=& R_A^{(2),ana} 
                      + R_A^{(2),rem},
\\
R_{\it NA}^{(2)}(s) &=& R_{\it NA}^{g,(2),ana} 
                      + R_{\it NA}^{g,(2),rem}.
\end{eqnarray}
$R^{(2),ana}(s)$ contains the analytically known 
threshold and high energy behaviour and is chosen in
such a way that $R^{(2),rem}(s)$ 
vanishes both 
at threshold and at the high energies. In the following 
the explicit form of $R^{(2),ana}$ is presented and the
numerical approximation of $R^{(2),rem}$ is discussed.

Let us first consider the $C_F^2$ part.
Taking the imaginary part of Eq.~(\ref{pofwa}) and solving 
formally for $R_A^{(2)}(s)$ we arrive at
\begin{eqnarray}
R_A^{(2)} + 4 R^{(1)} &=& \tilde{R}_A^{(2)}  - \hat{R}_A^{(2)}
\end{eqnarray}
$\hat{R}_A^{(2)}$ corresponds to the imaginary part of the
last term in Eq.~(\ref{pofwa}) and contains the constant and
$m^2/s$ terms of the l.h.s. in the limit $s\to\infty$;
$\tilde{R}_A^{(2)}$ starts at ${\cal O}(m^4/s^2)$.
Per construction $\tilde{R}_A^{(2)}$ has a $1/v$ singularity at threshold 
which should be subtracted in such way that the 
high energy behaviour is not disturbed. This is achieved with
an additional factor $(1-v^2)^4$ and leads to
\begin{eqnarray}
R_A^{(2)} &=& R_A^{(2),rem} + (1-v^2)^4 R_A^{(2),thr}
                                        - 4 R^{(1)} - \hat{R}_A^{(2)}.
\nonumber\\
          &=& R_A^{(2),rem} + \frac{(1-v^2)^4}{v}\frac{3\pi^4}{8}
                            - 4 R^{(1)}
\nonumber\\
&& 
                +v\frac{2619}{64}-v^3\frac{2061}{64}
                +\frac{81}{8}\left(1-v^2\right)\ln\frac{1-v}{1+v}
\end{eqnarray}
In Figure~\ref{figrema} 
$R_A^{(2),rem}(s)$ is plotted against $p=(1-v)/(1+v)$ for different 
Pad\'e approximations. The task is now to find an 
analytically simple formula which describes the behaviour
of this remainder as close as possible. 
A closer look into the result of the Pad\'e approximation 
shows that an expansion of
$R_A^{(2),rem}(s)$ for high energies ($p\to 0$) has the form:
$R_A^{(2),rem}(s)=p^{3/2} (d_0 + d_1 p + d_2 p^2 + \ldots)$.
The appearance of the half-integer exponents is a relic
of the method in particular of the variable $\omega$
introduced in Eq.~(\ref{omega}).
The $p^{3/2}$ term imitates the quartic mass correction which
also contain $\ln^2 m^2$ pieces in an excellent way as can be seen from
Figure~\ref{fighigh}. 
Thus we performed
a fit in the variable $p$
using Legendre polynomials $P_n(p)$
multiplied by a factor $p^{3/2}(1-p)$ which ensures the
zeros at the boundary.
In Figure~\ref{figrema} one can see that for $p\to 0$ a fit with a low
order polynomial would fail to deliver an acceptable result, so
in a first step a suitable (numerical small) term of ${\cal O}(p^{3/2})$
is added to get a curve which can easily be parametrised including terms 
up to $n=2$. The result reads:
\begin{eqnarray}
R_A^{(2),rem}(s) &=& -198\left(\frac{m^2}{s}\right)^{3/2} 
\left(v^4-2v^2\right)^6 
\nonumber\\
&&
+100 p^{3/2} (1-p) \left[
2.21 P_0(p)
-1.57 P_1(p)
+0.27 P_2(p)
\right],
\label{rarem}
\end{eqnarray}
and is also shown in Figure~\ref{figrema}.

For the non-Abelian part, $R_{\it NA}^{(2)}$, we 
exploit
that the gluonic double bubble is known exactly
(see Eqs.~(\ref{pismartnagldb}) and (\ref{pofwnagldb})).
Proceeding as above leads to the result:
\begin{eqnarray}
R_{\it NA}^{(2)} &=& R_{\it NA}^{g,(2),rem} + R_{g}^{(2)}\Big|_{\xi=4}
%%%\nonumber\\ 
%%%&& 
+ v\frac{351}{32} - v^3\frac{297}{32},
\end{eqnarray}
with
\begin{eqnarray}
R_{\it NA}^{g,(2),rem} &=& -18\left(\frac{m^2}{s}\right)^{3/2} 
\left(v^4-2v^2\right)^4
\nonumber\\
&&
+50 p^{3/2} (1-p) \left[
1.73 P_0(p)
-1.24 P_1(p)
+0.64 P_2(p)
\right].
\label{rnarem}
\end{eqnarray}

%%%%%%%%%%%%%%%%%%%%%%%%%%%%%%%%%%%%%%%%%%%%%%%%%%%%%%%%%%%%
%%%%%%%%%%%%%%%%%%%%%%%%%%%%%%%%%%%%%%%%%%%%%%%%%%%%%%%%%%%%

\setcounter{equation}{0}
\section{\label{appdouble}Exact Result for Double Bubble Contribution}

In this appendix we list for convenience of the reader the exact result
of the imaginary part of the fermionic and gluonic double bubble diagram.
Following \cite{CheHoaKueSteTeu96} $R_l^{(2)}$, respectively, $R_{g}^{(2)}$
may be written as ($x=l,g$)
\begin{eqnarray}
R_x^{(2)}(s) &=& -\frac{1}{3}\left(
               R_\infty^x \ln\frac{\mu^2}{s} - R^x_0
                             \right)R^{(1)}
                 + 3 R_\infty^x \delta^{(2)},
\end{eqnarray}
with $R_{0,\infty}^x$ being the moments in the $\overline{\mbox{MS}}$
scheme with the results:
\begin{eqnarray}
R_\infty^l & = &  1,
\nonumber \\
R_0^l & = & - \frac{5}{3}+\ln 4,
\nonumber\\
R_\infty^{g} 
& = & -\frac{5}{4} - \frac{3}{8}\xi,
\nonumber \\
R_0^{g} 
& = & \frac{31}{12}-\frac{3}{4}\xi+\frac{3}{16}\xi^2
                    +\left(-\frac{5}{4}-\frac{3}{8}\xi\right)\ln 4.
\end{eqnarray}
$R^{(1)}$ is defined in~(\ref{r1exact}),
whereas $\delta^{(2)}$ was originally calculated in \cite{HoaKueTeu95}
and reads:
\begin{eqnarray}
\lefteqn{
\delta^{(2)} \, = \,
- \,\frac{\left( 3 - {{v }^2} \right) \,
       \left( 1 + {{v }^2} \right) }{6}\,\times\,}\nonumber\,\\ 
 & & \mbox{}\,
     \qquad\,\bigg\{\,\mbox{Li}_3(p) - 2\,\mbox{Li}_3(1 - p) - 
       3\,\mbox{Li}_3({p^2}) - 4\,\mbox{Li}_3\Big({p\over {1 + p}}\Big) - 
       5\,\mbox{Li}_3(1 - {p^2}) + 
       \frac{11}{2}\,\zeta(3)\,\nonumber\,\\ 
 & & \mbox{}\,\qquad + 
       \mbox{Li}_2(p)\,\ln\Big(\frac{4\,\left( 1 - {{v }^2} \right) }{
          {{v }^4}}\Big) + 2\,\mbox{Li}_2({p^2})\,
        \ln\Big(\frac{1 - {{v }^2}}{2\,{{v }^2}}\Big) + 
       2\,\zeta(2)\,\bigg[\, \ln p - 
          \ln\Big(\frac{1 - {{v }^2}}{4\,v }\Big) \,\bigg] 
      \,\nonumber\,
        \\ 
 & & \mbox{}\,\qquad - 
       \frac{1}{6}\,\ln\Big(\frac{1 + v }{2}\Big)\,
        \bigg[\, 36\,\ln 2\,\ln p - 44\,\ln^2 p + 
          49\,\ln p\,\ln\Big(\frac{1 - {{v }^2}}{4}\Big) + 
          \ln^2\Big(\frac{1 - {{v }^2}}{4}\Big) \,\bigg] \,\nonumber\,
        \\ 
 & & \mbox{}\,\qquad - 
       \frac{1}{2}\,\ln p\,\ln v\,
        \bigg[\, 36\,\ln 2 + 21\,\ln p + 16\,\ln v  - 
          22\,\ln(1 - {{v }^2}) \,\bigg]  \,\bigg\} \,\nonumber\,
     \\ 
 & & \mbox{}   + 
  \frac{1}{24}\,\bigg\{ \,
      ( 15 - 6\,{{v }^2} - {{v }^4} ) \,
      \Big( \mbox{Li}_2(p) + \mbox{Li}_2({p^2}) \Big)  + 
     3\,( 7 - 22\,{{v }^2} + 7\,{{v }^4} ) \,
      \mbox{Li}_2(p)\,\nonumber\,\\ 
 & & \mbox{}\,\qquad - 
     ( 1 - v  ) \,
      ( 51 - 45\,v  - 27\,{{v }^2} + 5\,{{v }^3} ) \,
      \zeta(2)\,\nonumber\,\\[2mm] 
 & & \mbox{}\,\qquad + 
     \frac{\left( 1 + v  \right) \,
        \left( -9 + 33\,v  - 9\,{{v }^2} - 15\,{{v }^3} + 
          4\,{{v }^4} \right) }{v }\,\ln^2 p\,\nonumber\,
      \\ 
 & & \mbox{}\,\qquad + 
     \bigg[ \,( 33 + 22\,{{v }^2} - 7\,{{v }^4} ) \,
         \ln 2 - 10\,( 3 - {{v }^2} ) \,
         ( 1 + {{v }^2} ) \,\ln v \,
   \nonumber\,    \\ 
 & & \mbox{}\,\qquad\,\qquad\,\qquad   - 
        ( 15 - 22\,{{v }^2} + 3\,{{v }^4} ) \,
         \ln\Big(\frac{1 - {{v }^2}}{4\,{{v }^2}}\Big)\,\bigg] \,
      \ln p\,\nonumber\,\\ 
 & & \mbox{}\,\qquad + 
     2\,v \,( 3 - {{v }^2} ) \,
   \ln\Big(\frac{4\,\left( 1 - {{v }^2} \right) }{{{v }^4}}\Big)\,
   \bigg[\, \ln v - 3\,\ln\Big(\frac{1 - {{v }^2}}{4\,v }\Big) 
      \,\bigg]
       \,\nonumber\,\\ 
 & & \mbox{}\,\qquad + 
     \frac{237 - 96\,v  + 62\,{{v }^2} + 32\,{{v }^3} - 
        59\,{{v }^4}}{4}\,\ln p - 
     16\,v \,( 3 - {{v }^2} ) \,\ln\Big(\frac{1 + v }{4}\Big)\,
      \nonumber\,\\ 
 & & \mbox{}\,\qquad - 
     2\,v \,( 39 - 17\,{{v }^2} ) \,
      \ln\Big(\frac{1 - {{v }^2}}{2\,{{v }^2}}\Big) - 
     \frac{v \,\left( 75 - 29\,{{v }^2} \right) }{2}\,\bigg\} 
\,.
\end{eqnarray}

\end{appendix}

%%%%%%%%%%%%%%%%%%%%%%%%%%%%%%%%%%%%%%%%%%%%%%%%%%%%%%%%%%%%%%%%%%%%%%%%
\sloppy
\raggedright
\def\app#1#2#3{{\it Act. Phys. Pol. }{\bf B #1} (#2) #3}
\def\apa#1#2#3{{\it Act. Phys. Austr.}{\bf #1} (#2) #3}
\def\lhc{Proc. LHC Workshop, CERN 90-10}
\def\npb#1#2#3{{\it Nucl. Phys. }{\bf B #1} (#2) #3}
\def\plb#1#2#3{{\it Phys. Lett. }{\bf B #1} (#2) #3}
\def\prd#1#2#3{{\it Phys. Rev. }{\bf D #1} (#2) #3}
\def\pR#1#2#3{{\it Phys. Rev. }{\bf #1} (#2) #3}
\def\prl#1#2#3{{\it Phys. Rev. Lett. }{\bf #1} (#2) #3}
\def\prc#1#2#3{{\it Phys. Reports }{\bf #1} (#2) #3}
\def\cpc#1#2#3{{\it Comp. Phys. Commun. }{\bf #1} (#2) #3}
\def\nim#1#2#3{{\it Nucl. Inst. Meth. }{\bf #1} (#2) #3}
\def\pr#1#2#3{{\it Phys. Reports }{\bf #1} (#2) #3}
\def\sovnp#1#2#3{{\it Sov. J. Nucl. Phys. }{\bf #1} (#2) #3}
\def\jl#1#2#3{{\it JETP Lett. }{\bf #1} (#2) #3}
\def\jet#1#2#3{{\it JETP Lett. }{\bf #1} (#2) #3}
\def\zpc#1#2#3{{\it Z. Phys. }{\bf C #1} (#2) #3}
\def\ptp#1#2#3{{\it Prog.~Theor.~Phys.~}{\bf #1} (#2) #3}
\def\nca#1#2#3{{\it Nouvo~Cim.~}{\bf #1A} (#2) #3}
%%%%%%%%%%%%%%%%%%%%%%%%%%%%%%%%%%%%%%%%%%%%%%%%%%%%%%%%%%%%%%%%%%%%%%%%

%%%%%%%%%%%%%%%%%%%%%%%%%%%%%%%%%%%%%%%%%%%%%%%%%%%%%%%%%%%%
%%%%%%%%%%%%%%%%%%%%%%%%%%%%%%%%%%%%%%%%%%%%%%%%%%%%%%%%%%%%

\begin{figure}[ht]
 \begin{center}
 \begin{tabular}{c}
   \epsfxsize=11.5cm
   \leavevmode
   \epsffile[110 330 460 520]{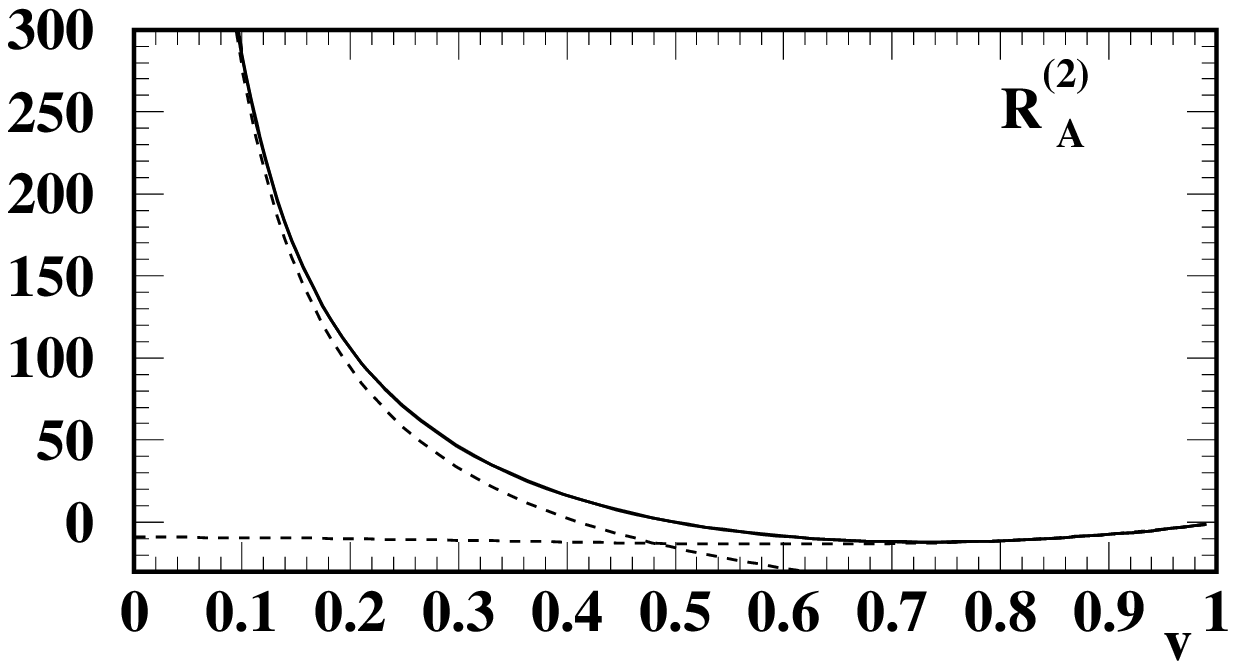}  
   \\
   \epsfxsize=11.5cm
   \leavevmode
   \epsffile[110 330 460 520]{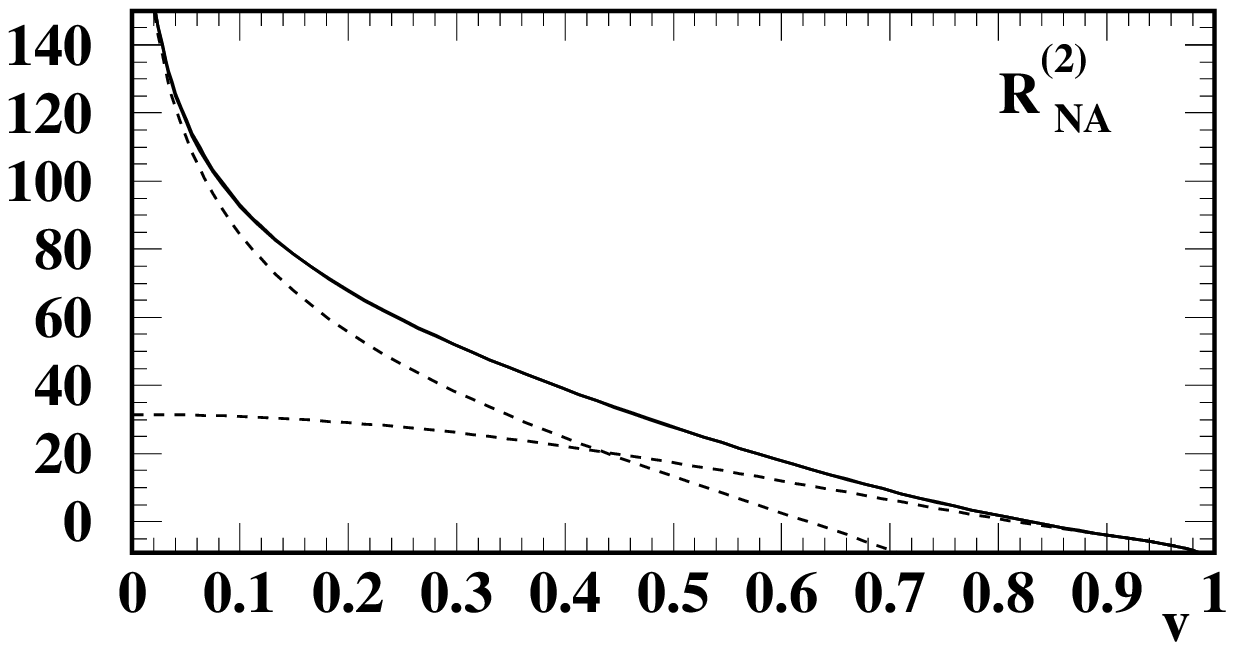} 
   \\
   \epsfxsize=11.5cm
   \leavevmode
   \epsffile[110 330 460 520]{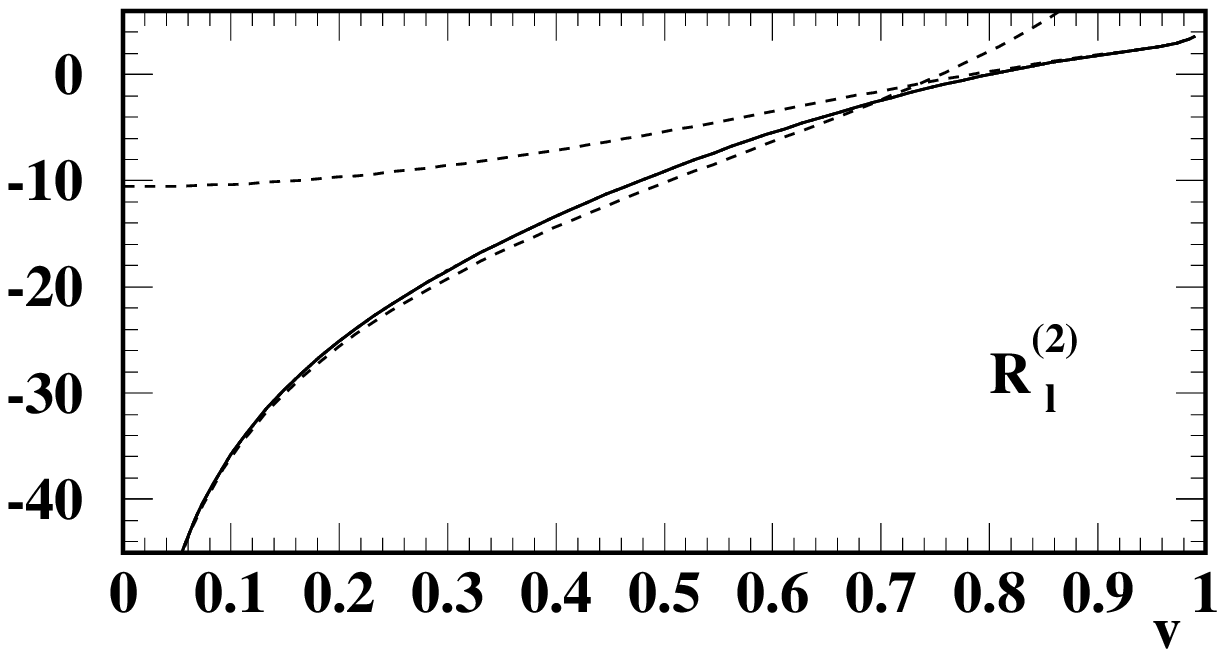}
 \end{tabular}
 \caption{\label{figfull} Complete results
                          plotted against
                          $v=\protect\sqrt{1-4m^2/s}$. The
                          high energy approximation includes the $m^4/s^2$ 
                          term.}
 \end{center}
\end{figure}

\begin{figure}[ht]
 \begin{center}
 \begin{tabular}{c}
   \epsfxsize=11.5cm
   \leavevmode
   \epsffile[110 330 460 520]{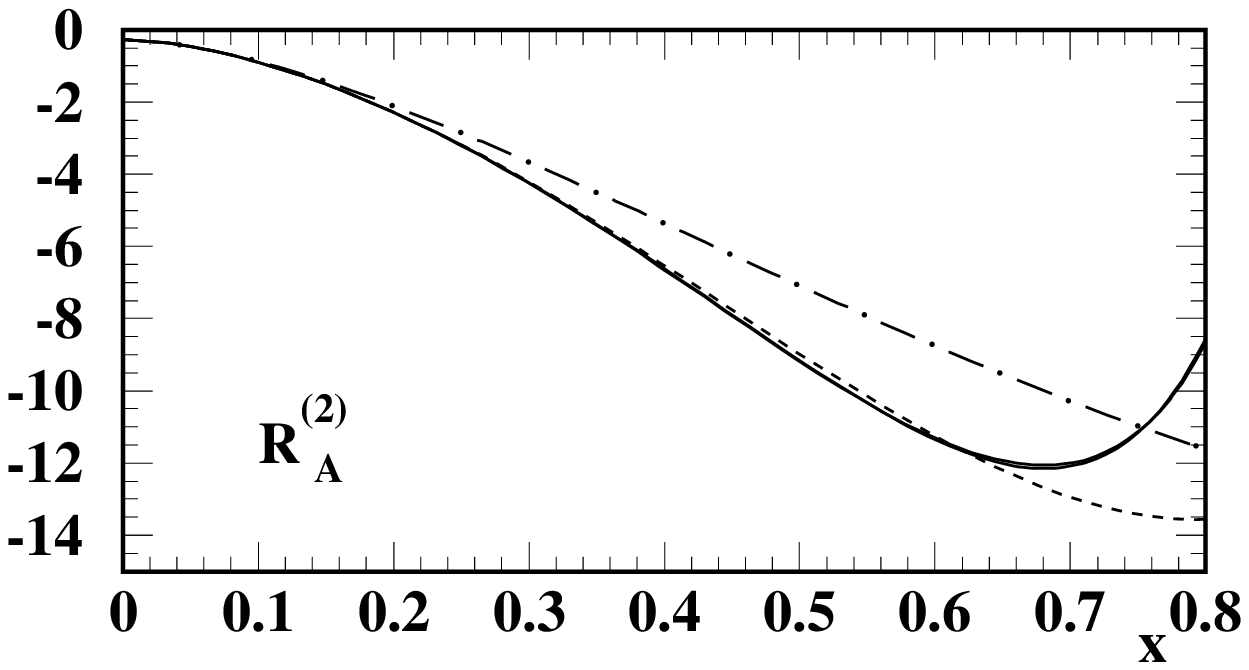}  
   \\
   \epsfxsize=11.5cm
   \leavevmode
   \epsffile[110 330 460 520]{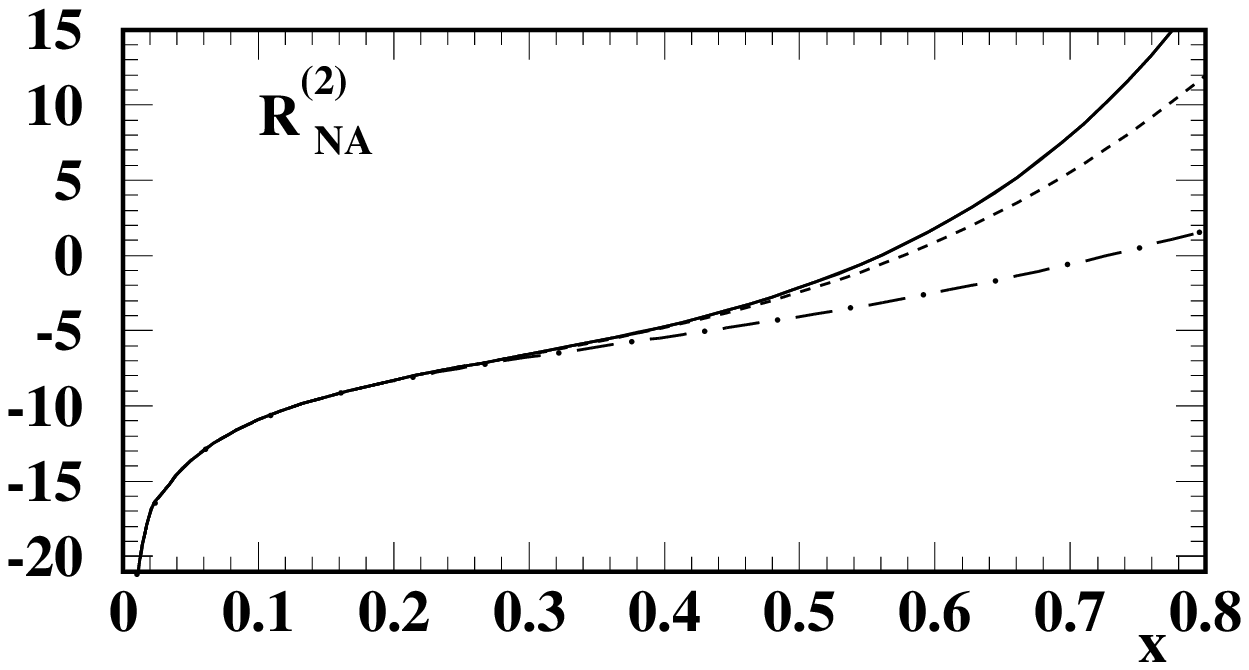} 
   \\
   \epsfxsize=11.5cm
   \leavevmode
   \epsffile[110 330 460 520]{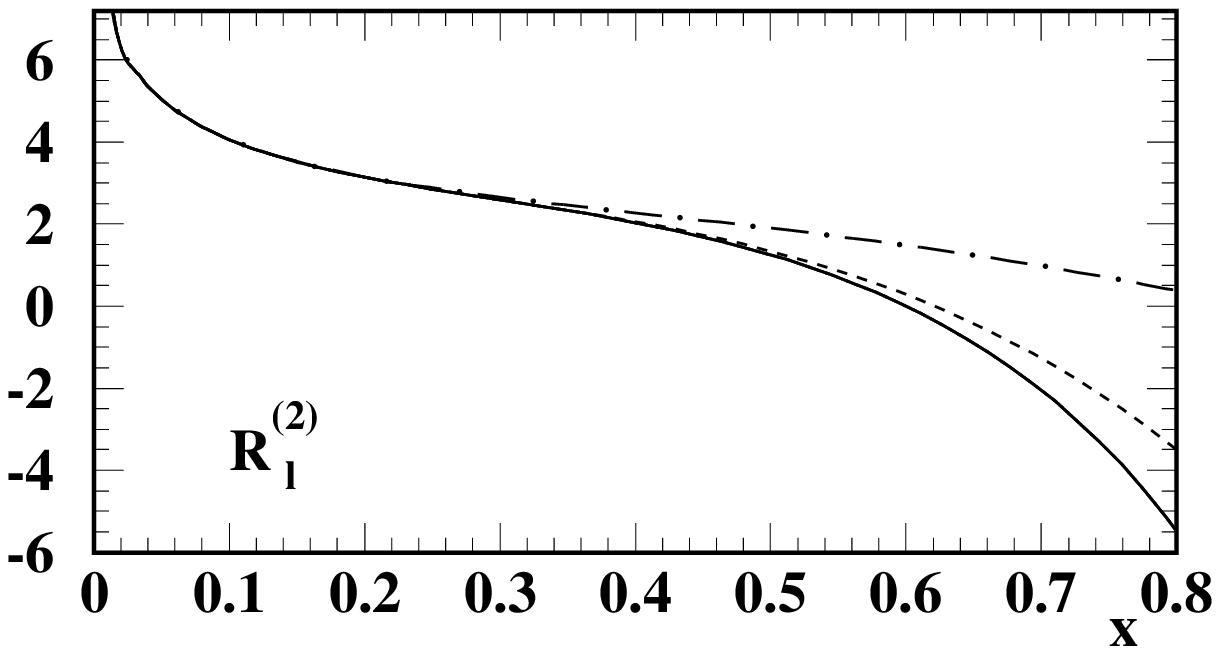} 
 \end{tabular}
 \caption{\label{fighigh}High energy region ($x=2m/\protect\sqrt{s}$). 
                         The complete results (full line) are
                         compared to the high energy
                         approximations including the $m^2/s$
                         (dash-dotted) and the $m^4/s^2$ (dashed) terms.}
 \end{center}
\end{figure}

\begin{figure}[ht]
 \begin{center}
 \begin{tabular}{c}
   \epsfxsize=11.5cm
   \leavevmode
   \epsffile[110 330 460 520]{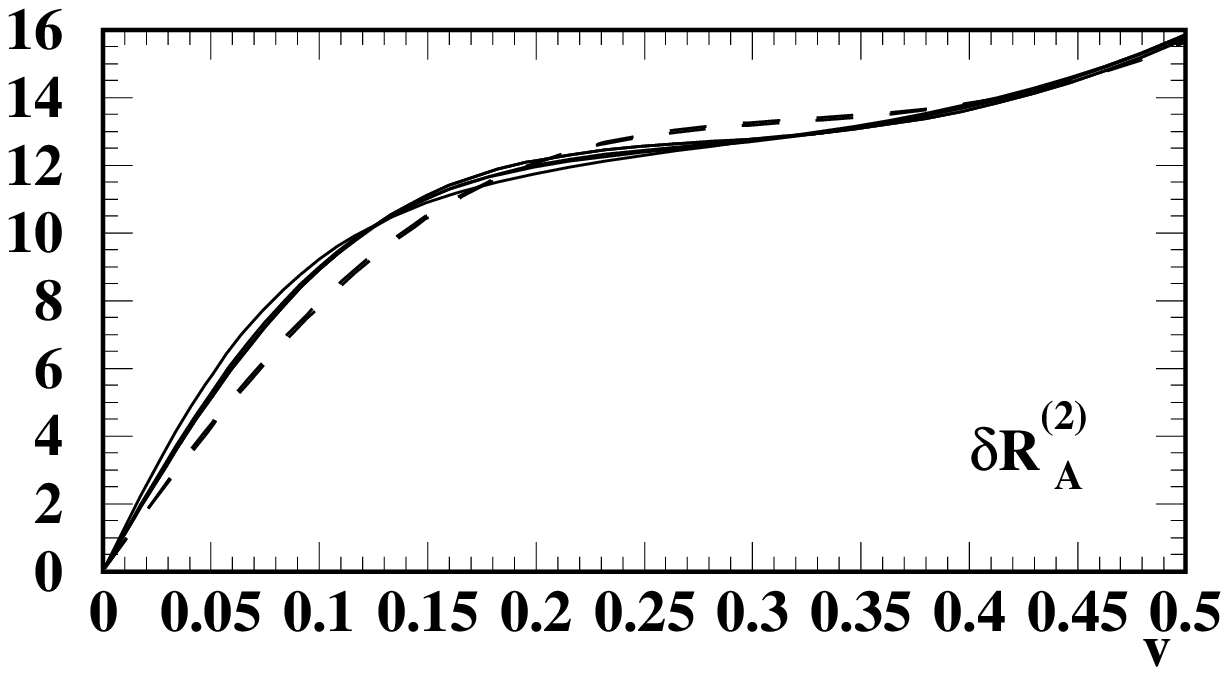}  
   \\
   \epsfxsize=11.5cm
   \leavevmode
   \epsffile[110 330 460 520]{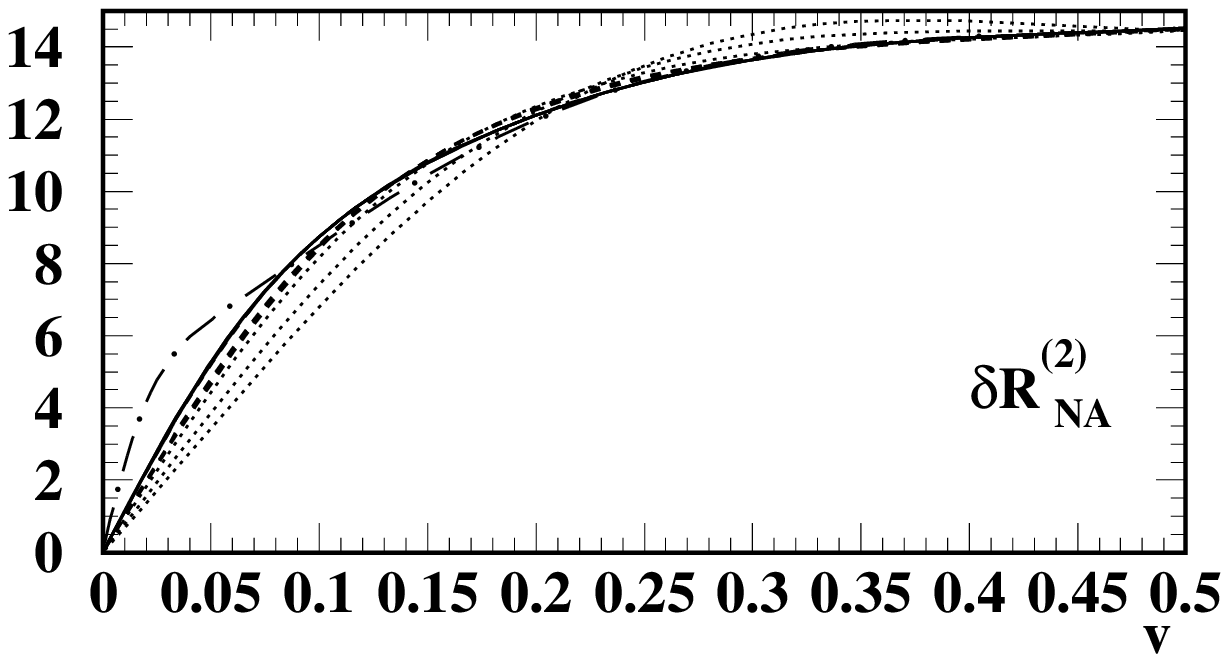} 
   \\
   \epsfxsize=11.5cm
   \leavevmode
   \epsffile[110 330 460 520]{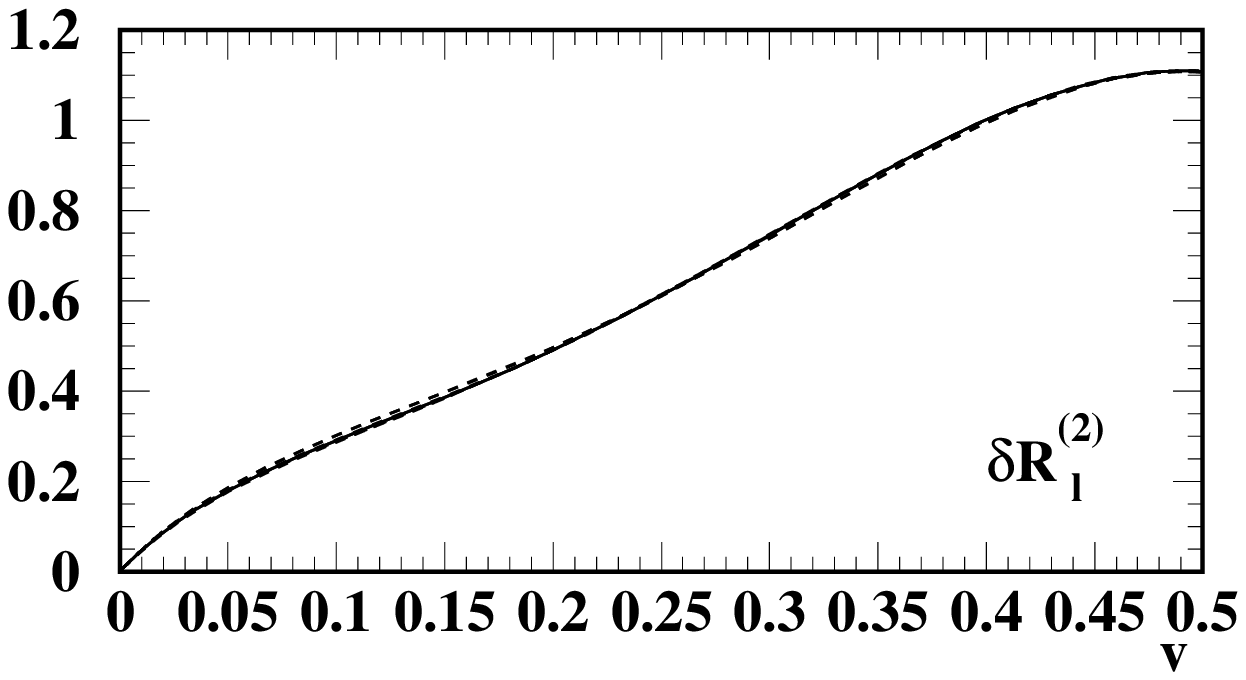} 
 \end{tabular}
 \caption{\label{figthr}Threshold behaviour of the remainder $\delta R^{(2)}$ 
                        for three different Pad\'e approximants. (The singular
                        and constant parts around threshold are subtracted.)} 
 \end{center}
\end{figure}

%%%%%%%%%%%%%%%%%%%%%%%%%%%%%%%%%%%%%%%%%%%%%%%%%%%%%%%%%%%%
%%%%%%%%%%%%%%%%%%%%%%%%%%%%%%%%%%%%%%%%%%%%%%%%%%%%%%%%%%%%

\begin{figure}[ht]
 \begin{center}
 \begin{tabular}{c}
   \epsfxsize=11.5cm
   \leavevmode
   \epsffile[110 330 460 520]{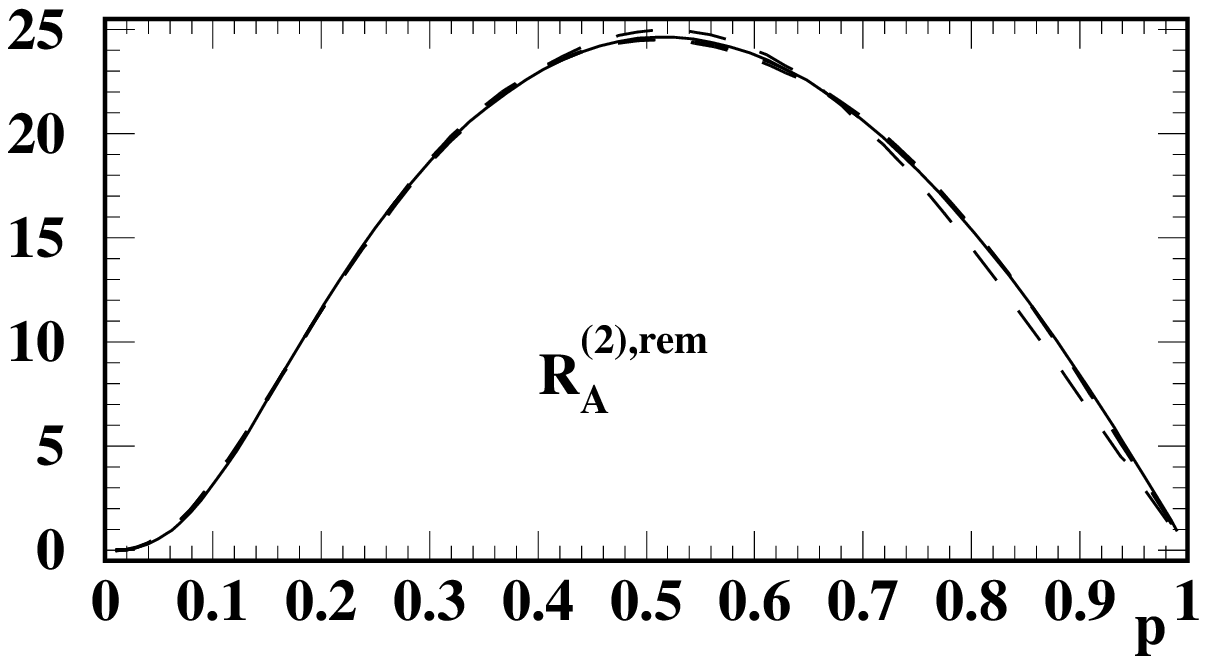}
 \end{tabular}
 \caption{\label{figrema} The remainder $R_A^{(2),rem}$ plotted for 
      different Pad\'e approximants (dashed lines) together with 
      the result of the fit (Eq.~(\ref{rarem}), solid line).}
 \end{center}
\end{figure}

\begin{figure}[ht]
 \begin{center}
 \begin{tabular}{c}
   \epsfxsize=11.5cm
   \leavevmode
   \epsffile[110 330 460 520]{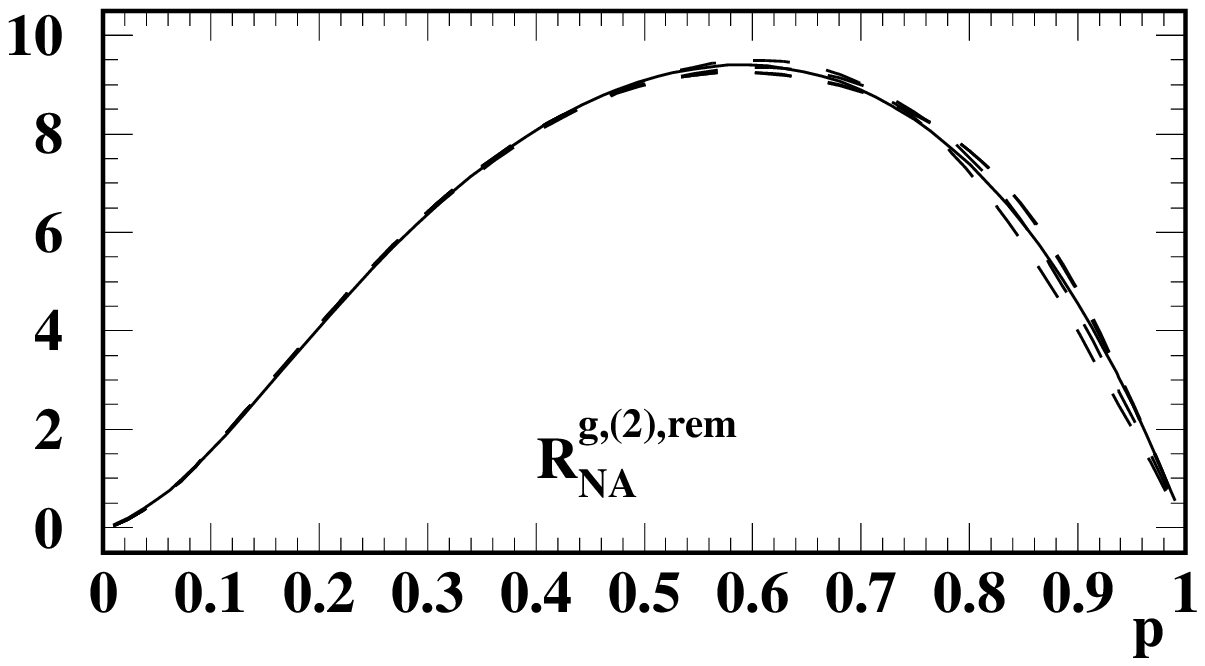}
 \end{tabular}
 \caption{\label{figremnaxi} The remainder 
      $R_{\it NA}^{g,(2),rem}$ plotted for 
      different Pad\'e approximants (dashed lines) together with 
      the result of the fit (Eq.~(\ref{rnarem}), solid line).}
 \end{center}
\end{figure}

\end{document}